\documentclass[final,hidelinks,onefignum,onetabnum]{siamart220329}

\usepackage{lipsum}
\usepackage{amsfonts}
\usepackage{graphicx}
\usepackage{epstopdf}
\usepackage{algorithmic}
\usepackage{amsmath}
\usepackage{amssymb}
\usepackage{mathrsfs}
\usepackage{cite}
\usepackage{fnpct}
\usepackage[nohyperlinks,nolist]{acronym}

\ifpdf
  \DeclareGraphicsExtensions{.eps,.pdf,.png,.jpg}
\else
  \DeclareGraphicsExtensions{.eps}
\fi

\newsiamremark{remark}{Remark}
\newsiamremark{hypothesis}{Hypothesis}
\crefname{hypothesis}{Hypothesis}{Hypotheses}
\newsiamthm{claim}{Claim}

\headers{Lossy Data Compression By Adaptive Mesh Coarsening}{N. Böing, J. Holke, C. Hergl et al.}

\title{Lossy Data Compression By Adaptive Mesh Coarsening\thanks{Submitted to the editors July 24, 2024.
\funding{Helmholtz Association of German Research Centres}}}

\author{Niklas Böing\footnote{C\lowercase{orresponding author. (\email{niklas.boeing@dlr.de})}}~\textsuperscript{,}\thanks{German Aerospace Center (DLR), Cologne, Germany}
\and Johannes Holke\footnotemark[3]
\and Chiara Hergl\footnotemark[3]
\and Luca Spataro\footnotemark[3]
\and Gregor Gassner\thanks{Department of Mathematics and Computer Science, University of Cologne, Cologne, Germany}
\and Achim Basermann\footnotemark[3]
}

\usepackage{tikz}
\usetikzlibrary[patterns]
\usetikzlibrary{shapes.geometric}
\usepackage{pgfplots}
\pgfplotsset{compat=1.18}

\usepackage{todonotes}
\usepackage{xcolor,colortbl}
\definecolor{turquoise}{rgb}{0.19, 0.84, 0.78}

\newcommand{\coarsenmapplain}{\mathscr{J}}

\definecolor{CustomYellow}{HTML}{F8E685}
\definecolor{CustomBlue}{HTML}{6DB3D6}
\definecolor{CustomGreen}{HTML}{CBDB96}
\definecolor{CustomRed}{HTML}{E26969}
\definecolor{CustomGray}{HTML}{C2C2C2}
\definecolor{CustomDark}{HTML}{000000}
\definecolor{CustomLight}{HTML}{C2C2C2}

\definecolor{CustomRedElem1}{HTML}{eb9696}
\definecolor{CustomRedElem2}{HTML}{e88787}
\definecolor{CustomRedElem3}{HTML}{e57878}
\definecolor{CustomRedElem4}{HTML}{e26969}
\definecolor{CustomRedElem5}{HTML}{cb5f5f}
\definecolor{CustomRedElem6}{HTML}{b55454}
\definecolor{CustomRedElem7}{HTML}{9e4a4a}

\definecolor{CustomBlueElem1}{HTML}{b6d9eb}
\definecolor{CustomBlueElem2}{HTML}{a7d1e6}
\definecolor{CustomBlueElem3}{HTML}{99cae2}
\definecolor{CustomBlueElem4}{HTML}{8ac2de}
\definecolor{CustomBlueElem5}{HTML}{7cbbda}
\definecolor{CustomBlueElem6}{HTML}{6db3d6}
\definecolor{CustomBlueElem7}{HTML}{62a1c1}
\definecolor{CustomBlueElem8}{HTML}{578fab}
\definecolor{CustomBlueElem9}{HTML}{4c7d96}
\definecolor{CustomBlueElem10}{HTML}{416b80}

\definecolor{CustomGreenElem1}{HTML}{e5edcb}
\definecolor{CustomGreenElem2}{HTML}{e0e9c0}
\definecolor{CustomGreenElem3}{HTML}{dbe6b6}
\definecolor{CustomGreenElem4}{HTML}{d5e2ab}
\definecolor{CustomGreenElem5}{HTML}{d0dfa1}
\definecolor{CustomGreenElem6}{HTML}{cbdb96}
\definecolor{CustomGreenElem7}{HTML}{b7c587}
\definecolor{CustomGreenElem8}{HTML}{a2af78}
\definecolor{CustomGreenElem9}{HTML}{8e9969}
\definecolor{CustomGreenElem10}{HTML}{7a835a}

\definecolor{Plot5}{HTML}{000000}
\definecolor{Plot2}{HTML}{E69F00}
\definecolor{Plot3}{HTML}{56B4E9}
\definecolor{Plot4}{HTML}{009E73}
\definecolor{Plot7}{HTML}{F0E442}
\definecolor{Plot6}{HTML}{0072B2}
\definecolor{Plot1}{HTML}{D55E00}

\begin{acronym}
\acro{amr}[AMR]{adaptive mesh refinement}
\acro{sfc}[SFC]{space-filling curve}
\acroplural{sfc}[SFCs]{space-filling curves}
\acro{hpc}[HPC]{high performance computing}
\end{acronym}

\ifpdf
\hypersetup{
  pdftitle={Title},
  pdfauthor={Authors}
}
\fi

\begin{document}

\maketitle

% REQUIRED
\begin{abstract}
Today's scientific simulations, for example in the high-performance exascale sector, produce huge amounts of data.
Due to limited I/O bandwidth and available storage space, there is the necessity to reduce scientific data of high performance computing applications.
Error-bounded lossy compression has been proven to be an effective approach tackling the trade-off between accuracy and storage space.
Within this work, we are exploring and discussing error-bounded lossy compression solely based on adaptive mesh refinement techniques.
This compression technique is not only easily integrated into existing adaptive mesh refinement applications but also suits as a general lossy compression approach for arbitrary data in form of multi-dimensional arrays, irrespective of the data type.
Moreover, these techniques permit the exclusion of regions of interest and even allows for nested error domains during the compression.
The described data compression technique is presented exemplary on ERA5 data.
\end{abstract}

% REQUIRED
\begin{keywords}
Lossy Compression, Adaptive Mesh Refinement, Earth System Modelling
\end{keywords}

% REQUIRED
\begin{MSCcodes}
68P20, 68P30
\end{MSCcodes}

\section{Introduction}

Climate modelling and numerical weather prediction are important tools to not only anticipate climate change, but also to improve the understanding of today's climate.
Especially, under the apprehension of rising extreme weather events, large-scale simulations will play a key role in precise prediction, forecasting and mitigation of these scenarios.

Climate models or more generally, Earth System Models (ESMs) imitating the dynamics and the interactions of the components of the Earth System (e.g. atmosphere, oceans, ice sheets, etc.), perform simulations mainly by solving partial differential equations on discretized domains utilizing numerical methods \cite{CESM2, ECHAM6, ICONESM, gmd-5-87-2012, MESSy2}.

The rapid increase in computational power - especially towards exascale computing - has led to increasingly complex models and higher spatial resolutions within the simulations \cite{8586949}.
For example, within the ``Earth System Chemistry Integrated Modelling'' initiative, approximately $2$ PB of climate data has been produced and stored at German Climate Computing Centre's supercomputer with chemistry-climate simulations of the EMAC model \cite{messydata}.

Within the NICAM studies on the K computer at RIKEN (Advanced Institute for Computational Science, Japan), global weather and climate simulations with ``sub 1-kilometre" horizontal resolution have been performed, yielding $6$ TB of data for a single snapshot of all 2D and 3D simulation variables \cite{satoh2017}.

Storage capacity and file I/O are becoming limiting factors, especially  for high-resolution and/or long-term simulations \cite{8287761, 5959140}.
Therefore, a growing need for compression techniques evolves in order to address these issues.
Moreover, storage reduction is not only beneficial in terms of performance, but also comes with economical savings \cite{Kunkel13, Kunkel16}.

Usually, it is distinguished between lossless and lossy compression methods.
Lossless compression approaches are capable of reconstructing the original data, in contrast to lossy methods which are discarding some information irreversible.

In general, lossy compression methods give better compression ratios by trading a loss in the data's accuracy for smaller file sizes.

The application domain for lossless and lossy compression schemes may vary.
For example, lossless compression approaches may be favored in terms of check-pointing - in order to restart a simulation run at given time steps - whereas lossy compression techniques are often preferred regarding the visualization and post-processing of data.

When performing lossy compression, the introduced inaccuracy should be controllable by the user in order to be widely applicable.
This leads us to the domain of error-bounded lossy compression.

There is a broad variety of error-bounded lossy compression methods, e.g. \cite{di2016fast, lindstrom2014fixed, Iverson2012, Lakshminarasimhan13} to name a few.
These compressors mainly offer absolute and/or relative error criteria which are not violated during the compression.

In this paper we are examining and evaluating an error-bounded lossy compression approach utilizing techniques from the field of \ac{amr} based on \acp{sfc} \cite{Holke18, Bader2013} with compliance to absolute and relative error bounds.

Especially within the context of ESMs, compression based on \ac{amr} offers great chances due to the strong coupling of the method to the underlying geospatial domain of the data.
For example, this approach easily copes with different error domains within the data during the compression.
Moreover, the technique is highly parallelizable favoring it's application within large simulation software.

The rest of the paper is structured as follows.
The succeeding section \ref{SecRelWork} collects related work within the field of lossy compression techniques - in particular with focus on \ac{amr}.
Afterwards, in section \ref{SecAMR} we are introducing the methodology and nomenclature of tree-based \ac{amr} with regards to \acp{sfc}.
Subsequently, we are describing the application of \ac{amr} techniques as an error-bounded lossy compression approach in chapter \ref{SecCMC} based on absolute and relative error criteria.
Thereafter, we are presenting some competitive compression results in comparison to different state of the art lossy compression methods on ERA5 data in chapter \ref{SecResults} and discuss the outcome in the following chapter \ref{SecDiscussion}.
At last, we conclude and assess the application of a our error-bounded lossy compressor in chapter \ref{SecConclusion}.

\section{Related Work}\label{SecRelWork}

There is a wide variety of compression techniques for large-scale scientific datasets.
Generally, it is distinguished between lossless and lossy compression schemes.
In this paper we will focus on lossy compression.

Two of the most well-known lossy compressors for floating point data are SZ \cite{di2016fast} and ZFP \cite{lindstrom2014fixed}.

The error-bounded SZ compressor developed by Di and Cappello \cite{di2016fast} is a best-fit curve-fitting model that attempts to approximate data points by either a \textit{preceding neighbor fitting}, a \textit{linear-curve fitting}, or a \textit{quadratic-curve fitting} based on the previous consecutive decompressed values.
If a data point can be predicted by one of the former fittings, it is substituted by a two-bit code denoting the fitting.
However, if none of the methods predict the data point compliant to a user-defined error bound, the data is labeled as \textit{unpredictable} and is stored separately after being compressed by analyzing it's binary representation.

Lindstrom presented the ZFP compressor \cite{lindstrom2014fixed} for multi-dimensional floating point data arrays.
The data array is divided into blocks of size $4^{\text{dim}}$ which are compressed independently and stored with a user-defined number of bits per block.
The approach consists of aligning the data values within a block to a common exponent, converting the values into a fixed-form representation, a decorrelation step by applying an orthogonal block transformation, and an embedded encoding of the transformation coefficients.
Since the compression is applied independently at the \textit{block-level}, the compressed data supports a fast decompression and random access at a \textit{block-level}.

Lakshminarasimhan et al. \cite{Lakshminarasimhan13} introduced a lossy compressor called ISABELA for in-situ processing of large-scale scientific data.
It is especially suited for data analysis and visualization.
ISABELA is based on sorting and cubic B-Spline fitting.
The compression is error-bounded and the decompression is completely local, allowing for random access within the compressed data.

TTHRESH is a lossy compression approach based on tensor decomposition which was developed by Ballester-Ripoll et al. \cite{tthresh}.
It is an error-bounded compression which achieves it's data reduction by compressing the flattened non-truncated core of the higher-order singular value decomposition (HOSVD).
The compression of a subset of the \textit{bit-planes}, chosen according to the error-threshold, of the flattened core is lossless and achieved by run-length encoding and arithmetic coding.
Besides the core, this compressor also compresses the factor matrices, resulting from the decomposition, yielding to high compression ratios.
An advantage of a tensor-based approach is that the compressed data supports linear operations without the need of prior decompression.

Ainsworth et al. \cite{MGARD1, MGARD2, MGARD3, MGARD4} have released a series of articles dealing with multilevel techniques for compression and reduction of scientific data.
Central to the series is a technique for multigrid adaptive reduction of data - called MGARD - based on multilevel decomposition.

Gong et al. \cite{Gong2022} extended the compression by imposing region-wise compression error bounds paired with a detection algorithm of critical regions which is inspired by mesh refinement techniques. 

Hovhannisyan et al. \cite{MultiRes2014} developed a multiresolution-based discontinuous Galerkin scheme in which data compression is applied by the means of a multiresolution analysis and hard thresholding.
The method operates on a hierarchy of nested grids for data given on a uniformly refined grid yielding high compression due to the hard thresholding.

The application of \ac{amr} within the field of image compression has been examined for example by Feischl and Hackl \cite{Feischl2023}.
Their adaptive JPEG compression utilizes the adaptive tree approximation algorithm \cite{Binev2004} which produces meshes (consisting of rectangles) based on approximating a given target function in a near optimal trade-off regarding accuracy and cost.
The image to be compressed resembles the target function and on each rectangular mesh element the classical JPEG quantization is applied.

Strafella and Chapon \cite{Strafella2022} developed a low-storage data model for \ac{amr} meshes called lightAMR.
The data model is extended by a lossless compression functionality for floating point data based on prediction functions.

A pre-processing workflow in order to enhance the performance of the lossy compressor SZ for AMR data has been portrayed by Wang et al. \cite{wang2023amric}.
Another pre-processing stratgey for \ac{amr} data has been laid out by Li et. al \cite{li2023lamp}.
They improved the compression ratios of \ac{amr} data and reduced the runtime overhead for building the AMR hierachy during the preconditioning step compared to other methods with their LAMP (Level-Associated Mapping-Based Preconditioning) framework.

Luo et al. \cite{zmesh} brought together concepts of \ac{amr} and compression techniques for floating point data.
In particular, a level reordering technique for \ac{amr} applications, such that the \textit{locality} of the data induced by \acp{sfc} (i.e. Morton- and Hilbert-Curve) is preserved even across adjacent elements of different refinement levels.
This results in \textit{smoother} and more compressible data sets enhancing the application of ZFP and SZ. 

Wang et al. \cite{wang2022tac} used a similar approach proposing an error-bounded lossy compression approach for 3D \ac{amr} data which was later extended in \cite{wang2023tac}.
The authors compressed each refinement level of the \ac{amr} data separately.
Depending on the ``density'' of the \ac{amr} level, different pre-processing strategies are applied, namely an optimized sparse tensor representation for low density data, an adaptive k-D tree approach for medium density data, and a ghost shell padding strategy for high density data.

The foundation for this work was laid out by Spataro, Bader and Holke \cite{Spataro22, SpataroMA} utilizing adaptive mesh refinement techniques as a compression tool for gridded data from ECHAM/MESSy Atmospheric Chemistry (EMAC) \cite{MESSy2, ECHAM5} simulations.
The geospatial data of chemical tracers were mapped onto an adaptive mesh which was coarsened alongside the data resulting in a lossy compression scheme based solely on \ac{amr}.
We have further developed the compression approach by adding error-bounded absolute and relative error criteria as well as different compression modes and region-wise error domains.

\section{Adaptive Mesh Refinement}\label{SecAMR}
\begin{figure}
    \centering
    \begin{tikzpicture}[scale=0.4]
        % Uniform Mesh
        \draw[step=2.0,black,thick] (-22,0) grid (-14,8);
        
        % Adaptive Mesh
		\draw[black, thick] (-10.5,0) -- (-2.5,0) -- (-2.5,8) -- (-10.5,8) -- cycle;
		\draw[black, thick] (-8.5,8) -- (-8.5,6);
		\draw[black, thick] (-6.5,8) -- (-6.5,6);
		\draw[black, thick] (-4.5,8) -- (-4.5,6);
		\draw[black, thick] (-4.5,6) -- (-2.5,6);
		\draw[black, thick] (-4.5,4) -- (-2.5,4);
		\draw[black, thick] (-4.5,2) -- (-2.5,2);
		\draw[black, thick] (-10.5,0) -- (-4.5,0) -- (-4.5,6) -- (-10.5,6) -- cycle;
		\draw[black, thick] (-10.5,0) -- (-6.5,0) -- (-6.5,4) -- (-10.5,4) -- cycle;
		\draw[black, thick] (-4.5,2) -- (-6.5,2) -- (-6.5,4) -- (-4.5,4) -- cycle;
		\draw[black, thick] (-8.5,4) -- (-6.5,4) -- (-6.5,6) -- (-8.5,6) -- cycle;
		\draw[black, thick] (-6.5,1) -- (-4.5,1);
		\draw[black, thick] (-6.5,5) -- (-4.5,5);
		\draw[black, thick] (-10.5,5) -- (-8.5,5);
		\draw[black, thick] (-9.5,6) -- (-9.5,4);
		\draw[black, thick] (-5.5,6) -- (-5.5,4);
		\draw[black, thick] (-5.5,2) -- (-5.5,0);
		\draw[black, thick] (-3.5,8) -- (-3.5,6);
		\draw[black, thick] (-4.5,7) -- (-2.5,7);
		\draw[black, thick] (-9.5,5.5) -- (-8.5,5.5);
		\draw[black, thick] (-9,6) -- (-9,5);
    \end{tikzpicture}
    \caption{A uniform mesh (left) consisting only of elements of the same size in comparison to an adaptive mesh (right) whose elements may vary in size resulting in different resolutions throughout the mesh.}
    \label{UniformAdaptiveMesh}
\end{figure}
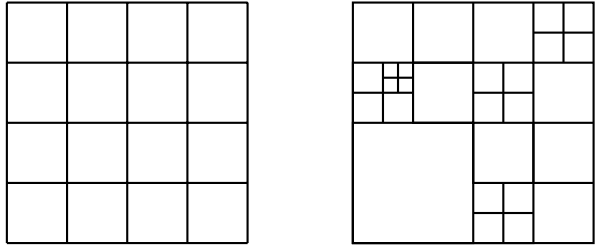

Meshes are often used by numerical solvers in order to discretize a computational domain, e.g. \cite{Braess_2007}.
There are different types of meshes being used within simulations.
Common are uniform meshes which have the same resolution throughout the domain.
Another type are \textit{adaptive meshes} consisting of areas of different resolutions; an exemplary comparison is shown in figure \ref{UniformAdaptiveMesh}.

Adaptive meshes allow to retain a fine resolution in certain areas of the mesh that require a small error and to establish a coarser resolution elsewhere.
The change within the resolution throughout the mesh may be dictated for example by local error estimators \cite{doi:10.1137/0715049, https://doi.org/10.1002/nme.1620240206}.

Although the handling of an adaptive mesh introduces some overhead, the amount of mesh elements can be significantly reduced resulting in faster runtimes than simulating to the same accuracy on a uniform mesh \cite{MULLER2013371, 10.1093/gji/ggs070}.
Adaptive meshes have been successfully used for decades in simulations and post-processing to reduce memory and compute time and increase simulation resolutions \cite{doi:10.1080/10618560310001634168, 10.1109/SC.2010.25}

Within this work we will look at \textit{tree-based adaptive meshes} being based on a \textit{coarse mesh} describing the topology of the domain \cite{Holke18, BursteddeWilcoxGhattas11, Peano}.

In particular we build our techniques on the \ac{amr} library \texttt{t8code} that implements tree-based \ac{amr} with space-filling curves for many different element shapes \cite{t8codev10, Holke_t8code_2023}.

These coarse mesh elements form the \textit{root elements} of \textit{refinement-trees}.
Each mesh element can be refined individually multiple times resulting in deeper hierarchies within the refinement-tree.
The refinement of a single element - a \textit{parent element} - results in several (finer) \textit{children elements}.

The \textit{refinement level} of an element corresponds to the depth in which it lays within the refinement tree. The root elements corresponds to refinement level zero and each recursive refinement of elements increases their refinement level.

Children elements which have the same refinement level and originated from the same parent element are called a \textit{family} of elements.

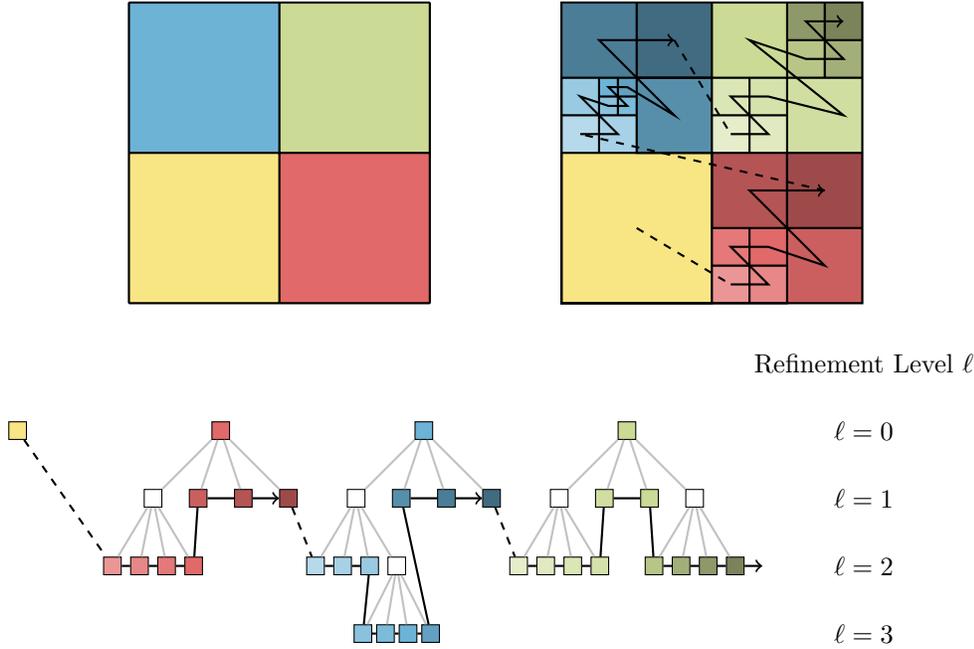
\begin{figure}
    \centering
    \begin{tikzpicture}[scale=0.5]
        % Uniform Mesh
        \draw[fill=CustomYellow] (-22,0) -- (-18,0) -- (-18,4) -- (-22,4) --cycle;
        \draw[fill=CustomRed] (-18,0) -- (-14,0) -- (-14,4) -- (-18,4) --cycle;
        \draw[fill=CustomBlue] (-22,4) -- (-18,4) -- (-18,8) -- (-22,8) --cycle;
        \draw[fill=CustomGreen] (-18,4) -- (-14,4) -- (-14,8) -- (-18,8) --cycle;
        \draw[step=4.0,black,thick,xshift=-2cm] (-20,0) grid (-12,8);
        
        % Adaptive Mesh
        \draw[fill=CustomYellow] (-10.5,0) -- (-6.5,0) -- (-6.5,4) -- (-10.5,4) --cycle;
        
        \draw[fill=CustomRedElem1] (-6.5,0) -- (-5.5,0) -- (-5.5,1) -- (-6.5,1) --cycle;
        \draw[fill=CustomRedElem2] (-5.5,0) -- (-4.5,0) -- (-4.5,1) -- (-5.5,1) --cycle;
        \draw[fill=CustomRedElem3] (-6.5,1) -- (-5.5,1) -- (-5.5,2) -- (-6.5,2) --cycle;
        \draw[fill=CustomRedElem4] (-5.5,1) -- (-4.5,1) -- (-4.5,2) -- (-5.5,2) --cycle;
        \draw[fill=CustomRedElem5] (-4.5,0) -- (-2.5,0) -- (-2.5,2) -- (-4.5,2) --cycle;
        \draw[fill=CustomRedElem6] (-6.5,2) -- (-4.5,2) -- (-4.5,4) -- (-6.5,4) --cycle;
        \draw[fill=CustomRedElem7] (-4.5,2) -- (-2.5,2) -- (-2.5,4) -- (-4.5,4) --cycle;
        
        \draw[fill=CustomBlueElem1] (-10.5,4) -- (-9.5,4) -- (-9.5,5) -- (-10.5,5) --cycle;
        \draw[fill=CustomBlueElem2] (-9.5,4) -- (-8.5,4) -- (-8.5,5) -- (-9.5,5) --cycle;
        \draw[fill=CustomBlueElem3] (-10.5,5) -- (-9.5,5) -- (-9.5,6) -- (-10.5,6) --cycle;
        \draw[fill=CustomBlueElem4] (-9.5,5) -- (-9,5) -- (-9,5.5) -- (-9.5,5.5) --cycle;
        \draw[fill=CustomBlueElem5] (-9,5) -- (-8.5,5) -- (-8.5,5.5) -- (-9,5.5) --cycle;
        \draw[fill=CustomBlueElem6] (-9.5,5.5) -- (-9,5.5) -- (-9,6) -- (-9.5,6) --cycle;
        \draw[fill=CustomBlueElem7] (-9,5.5) -- (-8.5,5.5) -- (-8.5,6) -- (-9,6) --cycle;
        \draw[fill=CustomBlueElem8] (-8.5,4) -- (-6.5,4) -- (-6.5,6) -- (-8.5,6) --cycle;
        \draw[fill=CustomBlueElem9] (-10.5,6) -- (-8.5,6) -- (-8.5,8) -- (-10.5,8) --cycle;
        \draw[fill=CustomBlueElem10] (-8.5,6) -- (-6.5,6) -- (-6.5,8) -- (-8.5,8) --cycle;
        
        \draw[fill=CustomGreenElem1] (-6.5,4) -- (-5.5,4) -- (-5.5,5) -- (-6.5,5) --cycle;
        \draw[fill=CustomGreenElem2] (-5.5,4) -- (-4.5,4) -- (-4.5,5) -- (-5.5,5) --cycle;
        \draw[fill=CustomGreenElem3] (-6.5,5) -- (-5.5,5) -- (-5.5,6) -- (-6.5,6) --cycle;
        \draw[fill=CustomGreenElem4] (-5.5,5) -- (-4.5,5) -- (-4.5,6) -- (-5.5,6) --cycle;
        \draw[fill=CustomGreenElem5] (-4.5,4) -- (-2.5,4) -- (-2.5,6) -- (-4.5,6) --cycle;
        \draw[fill=CustomGreenElem6] (-6.5,6) -- (-4.5,6) -- (-4.5,8) -- (-6.5,8) --cycle;
        \draw[fill=CustomGreenElem7] (-4.5,6) -- (-3.5,6) -- (-3.5,7) -- (-4.5,7) --cycle;
        \draw[fill=CustomGreenElem8] (-3.5,6) -- (-2.5,6) -- (-2.5,7) -- (-3.5,7) --cycle;
        \draw[fill=CustomGreenElem9] (-4.5,7) -- (-3.5,7) -- (-3.5,8) -- (-4.5,8) --cycle;
        \draw[fill=CustomGreenElem10] (-3.5,7) -- (-2.5,7) -- (-2.5,8) -- (-3.5,8) --cycle;

		\draw[black, thick] (-10.5,0) -- (-2.5,0) -- (-2.5,8) -- (-10.5,8) -- cycle;
		\draw[black, thick] (-8.5,8) -- (-8.5,6);
		\draw[black, thick] (-6.5,8) -- (-6.5,6);
		\draw[black, thick] (-4.5,8) -- (-4.5,6);
		\draw[black, thick] (-4.5,6) -- (-2.5,6);
		\draw[black, thick] (-4.5,4) -- (-2.5,4);
		\draw[black, thick] (-4.5,2) -- (-2.5,2);
		\draw[black, thick] (-10.5,0) -- (-4.5,0) -- (-4.5,6) -- (-10.5,6) -- cycle;
		\draw[black, thick] (-10.5,0) -- (-6.5,0) -- (-6.5,4) -- (-10.5,4) -- cycle;
		\draw[black, thick] (-4.5,2) -- (-6.5,2) -- (-6.5,4) -- (-4.5,4) -- cycle;
		\draw[black, thick] (-8.5,4) -- (-6.5,4) -- (-6.5,6) -- (-8.5,6) -- cycle;
		\draw[black, thick] (-6.5,1) -- (-4.5,1);
		\draw[black, thick] (-6.5,5) -- (-4.5,5);
		\draw[black, thick] (-10.5,5) -- (-8.5,5);
		\draw[black, thick] (-9.5,6) -- (-9.5,4);
		\draw[black, thick] (-5.5,6) -- (-5.5,4);
		\draw[black, thick] (-5.5,2) -- (-5.5,0);
		\draw[black, thick] (-3.5,8) -- (-3.5,6);
		\draw[black, thick] (-4.5,7) -- (-2.5,7);
		\draw[black, thick] (-9.5,5.5) -- (-8.5,5.5);
		\draw[black, thick] (-9,6) -- (-9,5);
		
		%SFC
		\draw[thick, CustomDark, dashed] (-8.5,2) -- (-6,0.5);
        \draw[thick, CustomDark, ->] (-6,0.5) -- (-5,0.5) -- (-6,1.5) -- (-5,1.5) -- (-3.5,1) -- (-5.5,3) -- (-3.5,3);
        \draw[thick, CustomDark, dashed] (-3.5,3) -- (-10,4.5);
        \draw[thick, CustomDark, ->] (-10,4.5) -- (-9,4.5) -- (-10,5.5) -- (-9.25,5.25) -- (-8.75,5.25) -- (-9.25,5.75) -- (-8.75,5.75) -- (-7.5,5) -- (-9.5,7) -- (-7.5,7);
        \draw[thick, CustomDark, dashed] (-7.5,7) -- (-6,4.5);
        \draw[thick, CustomDark, ->] (-6,4.5) -- (-5,4.5) -- (-6,5.5) -- (-5,5.5) -- (-3,5) -- (-5.5,7) -- (-4,6.5) -- (-3,6.5) -- (-4,7.5) -- (-3,7.5); 
        
    \end{tikzpicture}
    
    \vspace*{0.5cm}
    
    \begin{tikzpicture}[scale=0.3]
            %%Tree elements
            %Ebene 0
            \node (celem00) at (0,10) [rectangle, minimum width=0.2cm, fill=CustomYellow, draw]{};
            \node (celem01) at (9,10) [rectangle, minimum width=0.2cm, fill=CustomRed,draw]{};
            \node (celem02) at (18,10) [rectangle, minimum width=0.2cm, fill=CustomBlue,draw]{};
            \node (celem03) at (27,10) [rectangle, minimum width=0.2cm, fill=CustomGreen,draw]{};
            
            %Ebene 1
            \node (celem10) at (6,7) [rectangle, minimum width=0.2cm, draw]{};
            \node (celem11) at (8,7) [rectangle, minimum width=0.2cm, fill=CustomRedElem5, draw]{};
            \node (celem12) at (10,7) [rectangle, minimum width=0.2cm, fill=CustomRedElem6, draw]{};
            \node (celem13) at (12,7) [rectangle, minimum width=0.2cm, fill=CustomRedElem7, draw]{};
            
            \node (celem14) at (15,7) [rectangle, minimum width=0.2cm, draw]{};
            \node (celem15) at (17,7) [rectangle, minimum width=0.2cm, fill=CustomBlueElem8, draw]{};
            \node (celem16) at (19,7) [rectangle, minimum width=0.2cm, fill=CustomBlueElem9, draw]{};
            \node (celem17) at (21,7) [rectangle, minimum width=0.2cm, fill=CustomBlueElem10, draw]{};
            
            \node (celem18) at (24,7) [rectangle, minimum width=0.2cm, draw]{};
            \node (celem19) at (26,7) [rectangle, minimum width=0.2cm, fill=CustomGreenElem5, draw]{};
            \node (celem110) at (28,7) [rectangle, minimum width=0.2cm, fill=CustomGreenElem6, draw]{};
            \node (celem111) at (30,7) [rectangle, minimum width=0.2cm, draw]{};
            
            %Ebene 2
            \node (celem20) at (4.2,4) [rectangle, minimum width=0.2cm, fill=CustomRedElem1, draw]{};
            \node (celem21) at (5.4,4) [rectangle, minimum width=0.2cm, fill=CustomRedElem2, draw]{};
            \node (celem22) at (6.6,4) [rectangle, minimum width=0.2cm, fill=CustomRedElem3, draw]{};
            \node (celem23) at (7.8,4) [rectangle, minimum width=0.2cm, fill=CustomRedElem4, draw]{};
            
            \node (celem24) at (13.2,4) [rectangle, minimum width=0.2cm, fill=CustomBlueElem1, draw]{};
            \node (celem25) at (14.4,4) [rectangle, minimum width=0.2cm, fill=CustomBlueElem2, draw]{};
            \node (celem26) at (15.6,4) [rectangle, minimum width=0.2cm, fill=CustomBlueElem3, draw]{};
            \node (celem27) at (16.8,4) [rectangle, minimum width=0.2cm, draw]{};
			
			\node (celem28) at (22.2,4) [rectangle, minimum width=0.2cm, fill=CustomGreenElem1, draw]{};
            \node (celem29) at (23.4,4) [rectangle, minimum width=0.2cm, fill=CustomGreenElem2, draw]{};
            \node (celem210) at (24.6,4) [rectangle, minimum width=0.2cm, fill=CustomGreenElem3, draw]{};
            \node (celem211) at (25.8,4) [rectangle, minimum width=0.2cm, fill=CustomGreenElem4, draw]{};
            
            \node (celem212) at (28.2,4) [rectangle, minimum width=0.2cm, fill=CustomGreenElem7, draw]{};
            \node (celem213) at (29.4,4) [rectangle, minimum width=0.2cm, fill=CustomGreenElem8, draw]{};
            \node (celem214) at (30.6,4) [rectangle, minimum width=0.2cm, fill=CustomGreenElem9, draw]{};
            \node (celem215) at (31.8,4) [rectangle, minimum width=0.2cm, fill=CustomGreenElem10, draw]{};
            
            %Ebene 3
            \node (celem30) at (15.3,1) [rectangle, minimum width=0.2cm, fill=CustomBlueElem4, draw]{};
            \node (celem31) at (16.3,1) [rectangle, minimum width=0.2cm, fill=CustomBlueElem5, draw]{};
            \node (celem32) at (17.3,1) [rectangle, minimum width=0.2cm, fill=CustomBlueElem6, draw]{};
            \node (celem33) at (18.3,1) [rectangle, minimum width=0.2cm, fill=CustomBlueElem7, draw]{};
            
            \node (reflvltext) at (37.5,13) [draw=white] {Refinement Level $\ell$};
            \node (reflvl1) at (37.5,10) [draw=white] {$\ell = 0$};
            \node (reflvl2) at (37.5,7) [draw=white] {$\ell = 1$};
            \node (reflvl3) at (37.5,4) [draw=white] {$\ell = 2$};
            \node (reflvl4) at (37.5,1) [draw=white] {$\ell = 3$};

            %% Tree branches
            %Ebene 0-1
            \draw[CustomLight, thick] (celem01) -- (celem10);
            \draw[CustomLight, thick] (celem01) -- (celem11);
            \draw[CustomLight, thick] (celem01) -- (celem12);
            \draw[CustomLight, thick] (celem01) -- (celem13);
            
            \draw[CustomLight, thick] (celem02) -- (celem14);
            \draw[CustomLight, thick] (celem02) -- (celem15);
            \draw[CustomLight, thick] (celem02) -- (celem16);
            \draw[CustomLight, thick] (celem02) -- (celem17);
            
            \draw[CustomLight, thick] (celem03) -- (celem18);
            \draw[CustomLight, thick] (celem03) -- (celem19);
            \draw[CustomLight, thick] (celem03) -- (celem110);
            \draw[CustomLight, thick] (celem03) -- (celem111);
            
            %Ebene 1-2
            \draw[CustomLight, thick] (celem10) -- (celem20);
            \draw[CustomLight, thick] (celem10) -- (celem21);
            \draw[CustomLight, thick] (celem10) -- (celem22);
            \draw[CustomLight, thick] (celem10) -- (celem23);
            
            \draw[CustomLight, thick] (celem14) -- (celem24);
            \draw[CustomLight, thick] (celem14) -- (celem25);
            \draw[CustomLight, thick] (celem14) -- (celem26);
            \draw[CustomLight, thick] (celem14) -- (celem27);
            
            \draw[CustomLight, thick] (celem18) -- (celem28);
            \draw[CustomLight, thick] (celem18) -- (celem29);
            \draw[CustomLight, thick] (celem18) -- (celem210);
            \draw[CustomLight, thick] (celem18) -- (celem211);
            
            \draw[CustomLight, thick] (celem111) -- (celem212);
            \draw[CustomLight, thick] (celem111) -- (celem213);
            \draw[CustomLight, thick] (celem111) -- (celem214);
            \draw[CustomLight, thick] (celem111) -- (celem215);
            
            %Ebene 2-3
            \draw[CustomLight, thick] (celem27) -- (celem30);
            \draw[CustomLight, thick] (celem27) -- (celem31);
            \draw[CustomLight, thick] (celem27) -- (celem32);
            \draw[CustomLight, thick] (celem27) -- (celem33);
            
             %SFC
            \draw[thick, dashed, CustomDark] (celem00) -- (celem20);
            \draw[thick, ->, CustomDark] (celem20) -- (celem21) -- (celem22) -- (celem23) -- (celem11) -- (celem12) -- (celem13);
            \draw[thick, dashed, CustomDark] (celem13) -- (celem24);
            \draw[thick, ->, CustomDark] (celem24) -- (celem25) -- (celem26) -- (celem30) -- (celem31) -- (celem32) -- (celem33) -- (celem15) -- (celem16) -- (celem17);
            \draw[thick, dashed, CustomDark] (celem17) -- (celem28);
            \draw[thick, ->, CustomDark] (celem28) -- (celem29) -- (celem210) -- (celem211) -- (celem19) -- (celem110) -- (celem212) -- (celem213) -- (celem214) -- (celem215) -- (33,4);
    \end{tikzpicture}
    \caption{A coarse mesh (top left) consisting of the root elements of four trees (colored). The forest mesh resulting from different refinements of the four trees (top right). The underlying tree structure of the forest mesh is shown at the bottom. The root elements of the trees make up the refinement level zero. Recursive refinements increase the refinement level of the elements. The corresponding space filling curves of the trees indexing the forest mesh's elements are shown.}
    \label{AMRCmeshAndForest}
\end{figure}

The recursive refinements and the parent-child relation of elements result in a tree structure as shown in figure \ref{AMRCmeshAndForest}.
A new family of elements generated by a refinement covers the same geospatial domain as it's parent element.

The \textit{leaves} of a tree are the elements within the tree that are actually present within the adaptive mesh.
In particular, this corresponds to elements which do not have any children.

We denote a collection of trees as a \textit{forest}. Therefore, the adaptive mesh is referred to as a \textit{forest mesh}.

In figure \ref{AMRCmeshAndForest} we see the coarse mesh consisting of four elements (different in color) which were refined individually resulting in several tree-structures starting from the root element down to the leaves of the tree.

The level of the element in the graph gives information about the \textit{refinement level} of the element in the mesh.
Within adaptive meshes the levels of leaf elements can vary - which may result in non-conforming meshes.

One criterion for the refinement and the coarsening of elements is the local error estimation.
Areas with large local errors require a higher resolution than areas with a smaller local errors.

We use efficient \ac{sfc} techniques to assign a linear index to each element of a refinement tree thus ordering all elements in an array.
\acp{sfc} allow for fast computations of indices of elements and topological entities such as neighbor elements or parent elements~\cite{SFCOverview, Bader2013}.

\begin{figure}
    \centering
    \begin{tikzpicture}[scale=0.35]
        % Uniform Mesh 1
        \draw[step=4.0,black,thick] (0,0) grid (8,8);
        % SFC 1
        \draw[thick, CustomDark, ->] (2,2) -- (6,2) -- (2,6) -- (6,6);
        
        % Uniform Mesh 2
        \draw[step=2.0,black,thick] (10,0) grid (18,8);
        % SFC 2
        \draw[thick, CustomDark, ->] (11,1) -- (13,1) -- (11,3) -- (13,3) -- (15,1) -- (17,1) -- (15,3) -- (17,3) -- (11,5) -- (13,5) -- (11,7) -- (13,7) -- (15,5) -- (17,5) -- (15,7) -- (17,7);
        
        % Uniform Mesh 3
        \draw[step=1.0,black,thick] (20,0) grid (28,8);
        % SFC 3
        \draw[thick, CustomDark, ->] (20.5,0.5) -- (21.5,0.5) -- (20.5,1.5) -- (21.5,1.5) -- (22.5,0.5) -- (23.5,0.5) -- (22.5,1.5) -- (23.5,1.5) -- (20.5,2.5) -- (21.5,2.5) -- (20.5,3.5) -- (21.5,3.5) -- (22.5,2.5) -- (23.5,2.5) -- (22.5,3.5) -- (23.5,3.5) --
        (24.5,0.5) -- (25.5,0.5) -- (24.5,1.5) -- (25.5,1.5) -- (26.5,0.5) -- (27.5,0.5) -- (26.5,1.5) -- (27.5,1.5) -- (24.5,2.5) -- (25.5,2.5) -- (24.5,3.5) -- (25.5,3.5) -- (26.5,2.5) -- (27.5,2.5) -- (26.5,3.5) -- (27.5,3.5) --
        (20.5,4.5) -- (21.5,4.5) -- (20.5,5.5) -- (21.5,5.5) -- (22.5,4.5) -- (23.5,4.5) -- (22.5,5.5) -- (23.5,5.5) -- (20.5,6.5) -- (21.5,6.5) -- (20.5,7.5) -- (21.5,7.5) -- (22.5,6.5) -- (23.5,6.5) -- (22.5,7.5) -- (23.5,7.5) --
        (24.5,4.5) -- (25.5,4.5) -- (24.5,5.5) -- (25.5,5.5) -- (26.5,4.5) -- (27.5,4.5) -- (26.5,5.5) -- (27.5,5.5) -- (24.5,6.5) -- (25.5,6.5) -- (24.5,7.5) -- (25.5,7.5) -- (26.5,6.5) -- (27.5,6.5) -- (26.5,7.5) -- (27.5,7.5);
        
    \end{tikzpicture}
    \caption{The Z-Curve filling the unit square of different refinement levels.}
    \label{SFCMortonExample}
\end{figure}
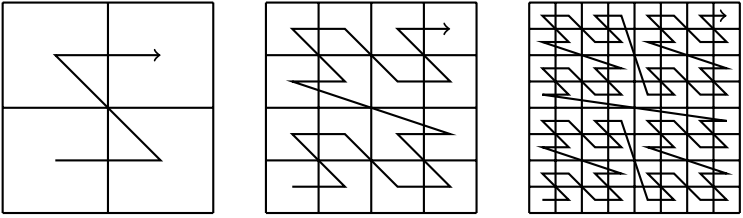

In this work, we are using the \textit{Morton-Curve} or \textit{Z-Curve} \cite{Morton1966} for quadrilateral and hexahedral elements as shown in figure \ref{SFCMortonExample}.
The indexing based on \acp{sfc} preserves the locality of the mesh elements regardless of their refinement levels. 
Moreover, the transformation from Cartesian reference coordinates to a Morton index (at a predefined refinement level) is computationally feasible by bit-wise interleaving of the coordinates.

\section{Approach}\label{SecCMC}

We introduce a lossy compression approach for geospatial data based on \ac{amr} techniques.
Within this section we are describing the compression based on adaptive coarsening compliant to controllable losses and the decompression of the data thereafter.

\subsection{Compression}
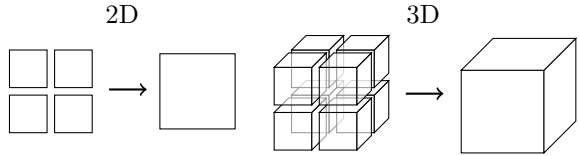
\begin{figure}
	\begin{center}
		\begin{tikzpicture}[scale=0.5]
			
			\node (twodlabel) at (-5.5, 2.15,0) {2D};
			\node (threedlabel) at (2.5, 2.15,0) {3D};
			%\draw[thick, dashed, gray] (-1.9,2.15) -- (-1.9,-1.5);
			
			% 2D quads
			\foreach \x in {-7.5,-6.3}{
				\foreach \y in {0,1.2}{
					\draw[black] (\x,\y,0) -- ++(-1,0,0) -- ++(0,-1,0) -- ++(1,0,0) -- ++(0,1,0) -- cycle;
				}
			}
			
			% Arrow to coarse quad
			\draw[thick, black, ->] (-5.85,0.15,0) -- (-4.85,0.15,0);
			
			% Coarse quad
			\draw[black] (-2.5,1.1,0) -- ++(-2,0,0) -- ++(0,-2,0) -- ++(2,0,0) -- ++(0,2,0) -- cycle;
			
			% 3D cubes
			\foreach \x in {0,1.2}{
				\foreach \y in {0,1.2}{
					\foreach \z in {0,1.2}{
						\draw[black,fill=white,fill opacity=0.6] (\x,\y,\z) -- ++(-1,0,0) -- ++(0,-1,0) -- ++(1,0,0) -- cycle;
						\draw[black,fill=white,fill opacity=0.6] (\x,\y,\z) -- ++(0,0,-1) -- ++(0,-1,0) -- ++(0,0,1) -- cycle;
						\draw[black,fill=white,fill opacity=0.6] (\x,\y,\z) -- ++(-1,0,0) -- ++(0,0,-1) -- ++(1,0,0) -- cycle;
					}
				}
			}
			
			% Arrow to coarsened cube
			\draw[thick, black, ->] (2.25,0.25,0.55) -- (3.25,0.25,0.55);
			
			% The corasened 3D Cube
			\draw[black,fill=white] (5.25,0.2,-1.2) -- ++(-2.2,0,0) -- ++(0,-2.2,0) -- ++(2.2,0,0) -- cycle;
			\draw[black,fill=white] (5.25,0.2,-1.2) -- ++(0,0,-2.2) -- ++(0,-2.2,0) -- ++(0,0,2.2) -- cycle;
			\draw[black,fill=white] (5.25,0.2,-1.2) -- ++(-2.2,0,0) -- ++(0,0,-2.2) -- ++(2.2,0,0) -- cycle;
			
		\end{tikzpicture}
	\end{center}
	\caption{Coarsening scheme of a family of elements in a 2D quadrilateral case (left) and a 3D hexahedral case (right). The default coarsening in 2D is $4:1$ and in 3D it is $8:1$.}
	\label{CoarseningSchemeQuadHex}
\end{figure}

In order to outline the compression based on \ac{amr} we are considering an array of multi-dimensional data of an arbitrary data type which needs to be compressed.
We are restricting ourselves to two- and three-dimensional geospatial data, although the concept extends to arbitrary dimensions and domain geometries.
Generally, the compression approach splits up in the following core routines:
\begin{itemize}
    \item[1.)] Construct an initial mesh.
    \item[2.)] Map each data point onto a single element.
    \item[3.)] Adaptive coarsening of the mesh and the data.
\end{itemize}
Therefore, we are obtaining the actual lossy compression by coarsening the data.
In particular, we are replacing several data points by an interpolated value - based on a predefined \ac{amr} coarsening/refinement scheme. 
This process is shown in figure~\ref{CoarseningSchemeQuadHex}.
 
We are constructing a base mesh consisting only of a single refinement tree onto which the data will be mapped, i.e. our forest mesh is based on a single coarse element.
Whether or not the data is defined on given coordinates or just a plain array without underlying geospatial dependencies, the refinement tree is constructed such that the data will reside within the domain of the tree.

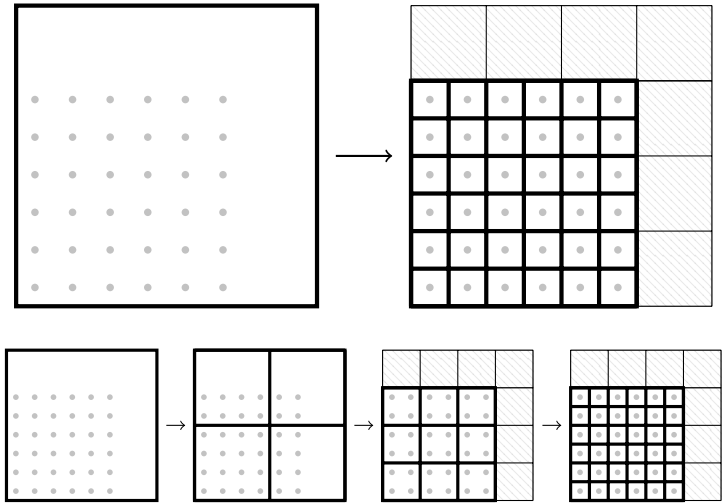
\begin{figure}
	\begin{center}
	\begin{tikzpicture}[scale=0.5]
		% Initial data points from the point cloud
		\foreach \x in {-10,-9,...,-5}{
			\foreach \y in {0.5,1.5,2.5,...,5.5} {
				\fill[CustomGray] (\x, \y) circle (0.1cm);
			}
		}
	
		\draw[black, ultra thick] (-10.5,0) -- (-2.5,0) -- (-2.5,8) -- (-10.5,8) -- cycle;
		
		% Arrow indicating the mapping
		\draw[black,thick,->] (-2,4) -- (-0.5,4);
		
		%\draw[pattern=north west lines, opacity=0.3] (0,5) -- (0,6) -- (6,6) -- (6,5) --cycle;
		\draw[pattern=north west lines, opacity=0.3] (0,6) -- (0,8) -- (8,8) -- (8,6) --cycle;
		\draw[pattern=north west lines, opacity=0.3] (6,0) -- (6,6) -- (8,6) -- (8,0) --cycle;
		
		% The grid
		\draw[step=1.0,black,thin] (0,5) grid (6,6);
		\draw[step=2.0,black,thin] (0,6) grid (8,8);
		\draw[step=2.0,black,thin] (6,0) grid (8,6);
		\draw[step=1.0,black, ultra thick] (0,0) grid (6,6);
		\draw[black, ultra thick] (0,0) -- (6,0) -- (6,6) -- (0,6) -- cycle;
		
		% Data points shown as element midpoints
		\foreach \x in {0.5,...,5.5}{
			\foreach \y in {0.5,...,5.5} {
				\fill[CustomGray] (\x, \y) circle (0.1cm);
			}
		}
	\end{tikzpicture}
	\vspace*{0.5cm}
	
	\begin{tikzpicture}[scale=0.5]
	    % Initial Mesh
	    \draw[black, very thick] (0,0) -- (4,0) -- (4,4) -- (0,4) -- cycle;
		% Initial data points
		\foreach \x in {0.25,0.75,...,2.75}{
			\foreach \y in {0.25,0.75,...,2.75} {
				\fill[CustomGray] (\x, \y) circle (0.075cm);
			}
		}
	    
	    % Arrow to second mesh
	    \draw[black, ->] (4.25,2) -- (4.75,2);
	    
	    % Second Mesh
	    \draw[step=2.0,black,very thick,xshift=-1cm] (6,0) grid (10,4);
	    \draw[black, very thick,xshift=-1cm] (6,0) -- (10,0) -- (10,4) -- (6,4) -- cycle;
	    \foreach \x in {5.25,5.75,...,7.75}{
			\foreach \y in {0.25,0.75,...,2.75} {
				\fill[CustomGray] (\x, \y) circle (0.075cm);
			}
		}
	    
	    % Arrow to third mesh
	    \draw[black, ->] (9.25,2) -- (9.75,2);
	    
	    % Third Mesh
	    \draw[pattern=north west lines, opacity=0.3] (10,3) -- (10,4) -- (14,4) -- (14,3) -- cycle;
	    \draw[pattern=north west lines, opacity=0.3] (13,0) -- (13,3) -- (14,3) -- (14,0) -- cycle;
	    \draw[step=1.0,black,thin] (10,0) grid (14,4);
	    \draw[step=1.0,black,very thick] (10,0) grid (13,3);
	    \draw[black, very thick] (10,0) -- (13,0) -- (13,3) -- (10,3) -- cycle;
	    \foreach \x in {10.25,10.75,...,12.75}{
			\foreach \y in {0.25,0.75,...,2.75} {
				\fill[CustomGray] (\x, \y) circle (0.075cm);
			}
		}
		
		 % Arrow to fourth mesh
	    \draw[black, ->] (14.25,2) -- (14.75,2);
	    
	    % Fourth Mesh
	    \draw[pattern=north west lines, opacity=0.3] (15,3) -- (15,4) -- (19,4) -- (19,3) -- cycle;
	    \draw[pattern=north west lines, opacity=0.3] (18,0) -- (18,3) -- (19,3) -- (19,0) -- cycle;
	    \draw[step=1.0,black,thin] (15,0) grid (19,4);
	    \draw[step=0.5,black,very thick] (15,0) grid (18,3);
	    \draw[black, very thick] (15,0) -- (18,0) -- (18,3) -- (15,3) -- cycle;
	    \foreach \x in {15.25,15.75,...,17.75}{
			\foreach \y in {0.25,0.75,...,2.75} {
				\fill[CustomGray] (\x, \y) circle (0.075cm);
			}
		}
		
	\end{tikzpicture}
	\caption{Exemplary domain of $6 \times 6$ data points embedded within a single quadrilateral refinement-tree (top left). The mesh based on the single refinement-tree is refined such that each data point is associated to a single element (top right). The ``dummy elements'' are shown as a hatched area. (Bottom:) An iterative construction of the initial mesh embedding the data ought to be compressed.}
	\label{InitialCompressionMesh}
	\end{center}
\end{figure}

Afterwards, the tree is refined until each data point is associated to a single element.
In case the cardinality of the data's dimensions are not identical and not a multiple of 2, and/or the underlying coordinates are non-equidistant, we may receive elements within the mesh that are not associated to an actual data point.
These so-called ``dummy''-elements are flagged with artificial ``missing values'' indicating the absence of data; see figure \ref{InitialCompressionMesh}.

Although we are potentially creating ``dummy elements'', this approach gives us more flexibility regarding the coarsening steps than modelling the data domain with several refinement trees.
In fact, it would be possible to coarsen all data points down to a single point (if the compression criterion allows for it) since the limit is the root element of the tree.

Essentially, this means that the underlying domain of the compressed data does not limit the adaptive coarsening process.
Since the ``dummy elements'' outside the underlying domain hold missing values, they are easily excluded from the interpolation yielding reasonable compression results at the domain boundaries.

In all of our experiments, dummy elements only make up a minor portion of all mesh elements, since they remain as coarse as possible.
Their contribution to the final compressed file size is negligible.
However, in most cases these elements and their missing values may even be discarded from the compressed data stream.

\begin{figure}
	\begin{center}
	\begin{tikzpicture}[scale=0.5]
	
	    %Axis
	    \draw[CustomGray, ->] (-0.5, -0.5) -- (8.5,-0.5);
	    \draw[CustomGray, ->] (-0.5, -0.5) -- (-0.5,8.5);
		
		%Area of dummy elements
	    \draw[pattern=north west lines, opacity=0.3] (0,6) -- (0,8) -- (8,8) -- (8,6) --cycle;
	    \draw[pattern=north west lines, opacity=0.3] (6,0) -- (6,6) -- (8,6) -- (8,0) --cycle;
		
		%Highlighted Area
		\draw[CustomBlue, fill=CustomBlue!40] (2,2) -- (4,2) -- (4,4) -- (2,4) -- cycle;
		% The grid
		\draw[step=1.0,black,thin] (0,5) grid (6,6);
		\draw[step=2.0,black,thin] (0,6) grid (8,8);
		\draw[step=2.0,black,thin] (6,0) grid (8,6);
		\draw[step=1.0,black, ultra thick] (0,0) grid (6,6);
		\draw[black, ultra thick] (0,0) -- (6,0) -- (6,6) -- (0,6) -- cycle;
		
		% Data points shown as element midpoints
		\foreach \x in {0.5,...,5.5}{
			\foreach \y in {0.5,...,5.5} {
				\fill[CustomGray] (\x, \y) circle (0.1cm);
			}
		}
		
		%SFC - Morton-Curve
		% The Space Filling Curve - i.e. the Morton Curve
		\draw[CustomRed, very thick, ->] (0.5,0.5) -- (1.5,0.5) -- (0.5,1.5) -- (1.5,1.5) -- (2.5,0.5) -- (3.5,0.5) -- (2.5,1.5) -- (3.5,1.5) -- (0.5,2.5) -- (1.5,2.5) -- (0.5,3.5) -- (1.5,3.5) -- (2.5,2.5) -- (3.5,2.5) -- (2.5,3.5) -- (3.5,3.5) -- (4.5,0.5) -- (5.5,0.5) -- (4.5,1.5) -- (5.5,1.5) -- (7,1) -- (4.5,2.5) -- (5.5,2.5) -- (4.5,3.5) -- (5.5,3.5) -- (7,3) -- (0.5,4.5) -- (1.5,4.5) -- (0.5,5.5) -- (1.5,5.5) -- (2.5,4.5) -- (3.5,4.5) -- (2.5,5.5) -- (3.5,5.5) -- (1,7) -- (3,7) -- (4.5,4.5) -- (5.5,4.5) -- (4.5,5.5) -- (5.5,5.5) -- (7,5) -- (5,7) -- (7,7);
		
		%Highlighted Area in Memory Layout
		\draw[CustomBlue, fill=CustomBlue!40] (4,-1.5) -- (6,-1.5) -- (6,-1) -- (4,-1) -- cycle;
		
		% First line of memory layout
		\draw[step=0.5,black] (-2,-1.5) grid (9,-1);
		\foreach \x in {-1.75,-1.25,...,8}
		{
			\fill[CustomGray] (\x, -1.25) circle (0.1cm);
		}
		\draw[pattern=north west lines, opacity=0.3] (8,-1.5) -- (8.5,-1.5) -- (8.5,-1) -- (8,-1) --cycle;
		\fill[CustomGray] (8.75, -1.25) circle (0.1cm);
		
		\draw (9.6,-1.25) node{$\ldots$};
		\draw (-1.25,-2.25) node{$\ldots$};
		
		% Second line of memory layout
		\draw[step=0.5,black] (-0.5,-2.5) grid (10,-2);
		
		\fill[CustomGray] (-0.25, -2.25) circle (0.1cm);
		\fill[CustomGray] (0.25, -2.25) circle (0.1cm);
		\fill[CustomGray] (0.75, -2.25) circle (0.1cm);
		\draw[pattern=north west lines, opacity=0.3] (1,-2.5) -- (1.5,-2.5) -- (1.5,-2) -- (1,-2) --cycle;
		\foreach \x in {1.75,2.25,...,5.25}
		{
			\fill[CustomGray] (\x, -2.25) circle (0.1cm);
		}
		\draw[pattern=north west lines, opacity=0.3] (5.5,-2.5) -- (6.5,-2.5) -- (6.5,-2) -- (5.5,-2) --cycle;
		\foreach \x in {6.75,7.25,...,8.25}
		{
			\fill[CustomGray] (\x, -2.25) circle (0.1cm);
		}
		\draw[pattern=north west lines, opacity=0.3] (8.5,-2.5) -- (10,-2.5) -- (10,-2) -- (8.5,-2) --cycle;

		%SFC - Memory Layout
		\draw[CustomRed, thick] (-1.75,-1.25) -- (9.15,-1.25);
		\draw[CustomRed, thick, ->] (-0.75,-2.25) -- (9.75,-2.25);
	\end{tikzpicture}
	\caption{The Morton ordering of the elements and the data respectively. The space-filling curve indexing the elements is shown in red. The locality of the elements in the ``physical domain'' is kept within the computational memory layout by this ordering (bottom). The highlighted elements (blue) indicate this correspondence.}
	\label{MortonCurve}
	\end{center}
\end{figure}
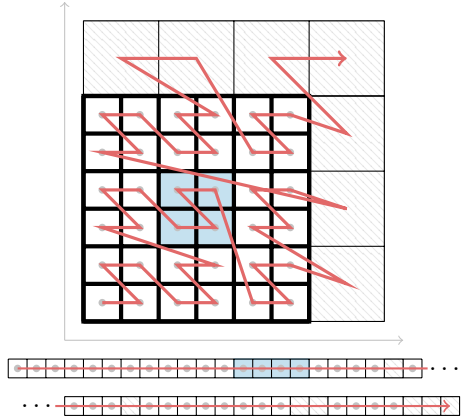

The elements of our tree-based scheme are indexed and ordered compliant to the Morton-Curve.
This corresponds to a linearization of the data, such that data which is close in the ``physical domain'' is also close in computational memory as indicated in figure \ref{MortonCurve}.

The ordering induced by the \ac{sfc} may even be beneficial if the lossy \ac{amr} compression is combined with additional lossless or lossy compression schemes afterwards.
For example techniques which are implemented trying to find redundancies in sliding windows like the Lempel-Ziv-Algorithm \cite{LZ77} - under the assumption that data which is spatially close is similar.

Performing lossy compression should be controllable, therefore, pre-defined data losses should be adjustable and not violated during the compression.

In the upcoming sections, we are presenting an absolute and a relative error criterion which ensure controllable losses during the compression based on \ac{amr}.
In particular, the criteria guarantee that the decompressed data does not deviate point-wise by more than the specified error threshold from the original data.

\subsubsection{Coarsening operation}

First, let us introduce the operation to coarsen a single family of elements.

We consider the situation where $n$ elements with associated data form a family that can be coarsened.
To simplify our notation, we assume that each element stores exactly one data point.
This assumption may be relaxed to an arbitrary number of points per element if required.

Let $\{e_i | i\in I\}$ be a family of $n=|I|$ elements with associated data points $\{x_i | i\in I\}$.
For the Morton curve, a family contains four members in 2D (quadrilaterals) and eight members in 3D (hexahedrons).

 When a family is coarsened into their parent element $\Bar{e}$, we calculate a new data value $\Bar{x}$ on $\Bar{e}$ that is interpolated from the $x_i$ data values which are associated to each of the children elements $\{e_i | i\in I\}$:
 
 \begin{equation}
     \coarsenmapplain\colon\left\{
     \begin{matrix}
     \mathbb{R}^n & \rightarrow & \mathbb{R} \\
 \{x_i | i\in I\} & \mapsto & \Bar{x}
     \end{matrix}
     \right.
 \end{equation}

There are different possible choices for the interpolation map
$\coarsenmapplain$.
For our first experiments we use the arithmetic mean as the interpolation

\begin{equation}
     \coarsenmapplain_{am}\colon\left\{
     \begin{matrix}
     \mathbb{R}^n & \rightarrow & \mathbb{R} \\
      \{x_i | i\in I\} & \mapsto & \frac{\sum_{i\in I} x_i}{n}.
     \end{matrix}
     \right.
     \label{ArithmeticMeanInterpolation}
 \end{equation}

\begin{figure}
	\begin{center}
		\begin{tikzpicture}[scale=0.7]
		
		    % The grid
			\draw[step=1.0,black,thick] (0,0) grid (4,4);
			\draw[black, thick] (0,0) -- (4,0) -- (4,4) -- (0,4) -- cycle;
			
			% The Space Filling Curve indices
			\draw (0.5,0.5) node{$x_{0}$};
			\draw (1.5,0.5) node{$x_{1}$};
			\draw (0.5,1.5) node{$x_{2}$};
			\draw (1.5,1.5) node{$x_{3}$};
			\draw (2.5,0.5) node{$x_{4}$};
			\draw (3.5,0.5) node{$x_{5}$};
			\draw (2.5,1.5) node{$x_{6}$};
			\draw (3.5,1.5) node{$x_{7}$};
			\draw (0.5,2.5) node{$x_{8}$};
			\draw (1.5,2.5) node{$x_{9}$};
			\draw (0.5,3.5) node{$x_{10}$};
			\draw (1.5,3.5) node{$x_{11}$};
			\draw (2.5,2.5) node{$x_{12}$};
			\draw (3.5,2.5) node{$x_{13}$};
			\draw (2.5,3.5) node{$x_{14}$};
			\draw (3.5,3.5) node{$x_{15}$};
			
			%Arrow
			\draw[black, ->] (4.5,2) -- (5.5,2);
			
			% The grid of the first coarsening
			\draw[step=2.0,black,thick] (6,0) grid (10,4);
			\draw[black, thick] (6,0) -- (10,0) -- (10,4) -- (6,4) -- cycle;
			
			\draw (7,1) node{$\Bar{x}_{0}$};
			\draw (9,1) node{$\Bar{x}_{1}$};
			\draw (7,3) node{$\Bar{x}_{2}$};
			\draw (9,3) node{$\Bar{x}_{3}$};
			
			%Arrow
			\draw[black, ->] (10.5,2) -- (11.5,2);
			
			% The grid of the second coarsening
			\draw[black, thick] (12,0) -- (16,0) -- (16,4) -- (12,4) -- cycle;
			
			\draw (14,2) node{$\Bar{\Bar{x}}_{0}$};
			
		\end{tikzpicture}
		\caption{(Left:) Data on the initial refinement level. (Middle:) A coarsening step in which all (four) element families have been coarsened. (Right:) A second coarsening step of the resulting family of elements from the first step.}
	\end{center}
	\label{TwoCoarseningsFromInitLevel}
\end{figure}
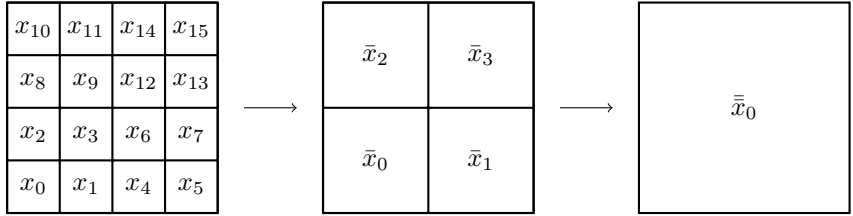

In the following subsections we are describing the adaptation criteria for the compression.
For portraying the criteria, we are assuming that any family $\left\{e_i | i \in I\right\}$ which is coarsened interpolates it's corresponding element-wise data $\left\{x_i | i \in I\right\}$ to $\Bar{x}$ by the means of the interpolation map \eqref{ArithmeticMeanInterpolation}, i.e. $\Bar{x}\, := \, \coarsenmapplain_{am}\left(\left\{x_i | i \in I\right\}\right)$.
Within figure~\ref{TwoCoarseningsFromInitLevel}, there are several element families displayed with their associated element-wise data as well as two succeeding coarsening iterations.

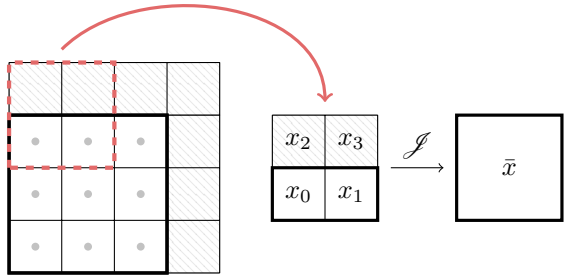
\begin{figure}
	\begin{center}
	\begin{tikzpicture}[scale=0.7]
	    
	    % Third Mesh
	    \draw[pattern=north west lines, opacity=0.3] (10,3) -- (10,4) -- (14,4) -- (14,3) -- cycle;
	    \draw[pattern=north west lines, opacity=0.3] (13,0) -- (13,3) -- (14,3) -- (14,0) -- cycle;
	    \draw[step=1.0,black,thin] (10,0) grid (14,4);
	    %\draw[step=1.0,black,very thick] (10,0) grid (13,3);
	    \draw[very thick] (10,0) -- (10,3) -- (13,3) -- (13,0) -- cycle;
	    \draw[black, very thick] (10,0) -- (13,0) -- (13,3) -- (10,3) -- cycle;
	    \foreach \x in {10.5,11.5,12.5}{
			\foreach \y in {0.5,1.5,2.5} {
				\fill[CustomGray] (\x, \y) circle (0.075cm);
			}
		}
		
		\draw [draw=CustomRedElem4, dashed, ultra thick] (10,2) -- (10,4) -- (12,4) -- (12,2) -- cycle; 
		\draw[CustomRedElem4,very thick,->] (11,4.25) to[out=45,in=90] (16,3.25);
		
		\draw[step=1.0,black,thin] (15,1) grid (17,3);
		\draw[pattern=north west lines, opacity=0.3] (15,2) -- (15,3) -- (17,3) -- (17,2) -- cycle;
		\draw [step=1.0,black,very thick] (15,1) -- (15,2) -- (17,2) -- (17,1) --cycle;
		\draw (15.5,1.5) node{$x_{0}$};
		\draw (16.5,1.5) node{$x_{1}$};
		\draw (15.5,2.5) node{$x_{2}$};
		\draw (16.5,2.5) node{$x_{3}$};
		
		\draw[black, ->] (17.25,2) -- node[midway,above](){$\coarsenmapplain$} (18.25,2);
		
		\draw[black, very thick] (18.5,1) -- (18.5,3) -- (20.5,3) -- (20.5,1) -- cycle;
		\draw (19.5,2) node{$\Bar{x}$};
		
	\end{tikzpicture}
	\caption{The boundary of the data domain (bold black) does not limit the adaptive coarsening if the interpolation ignores the ``missing values'' (corresponding to the hatched mesh elements). In this case, the interpolated value has to be computed only from $x_{0}$ and $x_{1}$.}
	\label{CompressBeyondDataBoundary}
	\end{center}
\end{figure}

We described that the domain of the variables will be embedded within an adaptive mesh potentially creating dummy elements which are flagged with missing values.
As mentioned before, the domain boundary of the data being compressed is not limiting the adaptive coarsening process.
In case the interpolation map is flexible enough to work with a varying amount of input values - ignoring elements that are flagged with missing values - the compression behaviour is well-defined.
An exemplary case is shown in figure \ref{CompressBeyondDataBoundary}.

During the compression we perform as much coarsening iterations as possible until we cannot coarsen the data any further without violating the user-specified errors.
One coarsening iteration plugs each family of elements (currently present in the mesh; i.e. leaf elements) into the interpolation functional and checks whether the interpolated value complies to the specified error bounds with regard to the initial data.
Per coarsening iteration, we may receive leaf elements with a refinement level of $\ell -1$.
The calculated inaccuracy from the interpolated value to the initial data - which results from the error compliance check - is associated to the coarse element, such that each element tracks its current maximum inaccuracy to all underlying initial data points.

In order to lower the memory demand and prevent load-imbalances in the parallelization, we do not hold the initial data in memory for the whole time of the compression to compare the interpolated coarse element values against each initial data point to check whether the coarsenings comply to the error bounds.

We only keep the (coarse) data defined on our current mesh after each coarsening iteration.
This implies that the calculation of the introduced inaccuracies during the first coarsening iteration is precise, since we still have the initial data for comparison.
However, in each subsequent coarsening iteration we do not have access to the initial data anymore and need to estimate the maximum inaccuracy of the upcoming coarsening to each underlying initial data point by the previous (coarse) data points of the family, the newly interpolated value for this family and the previous maximum inaccuracies to the initial data attached to each element.

Therefore, we are displaying the error calculation during the first coarsening iteration and the error estimation with regard to the initial data for each subsequent coarsening iterations for both error criteria in the following subsections. 

\subsubsection{Absolute error bounds}\label{LabAbsErrBounds}

When using absolute error bounds, the user pre-defines an absolute error value $\varepsilon_{max}$ (that may vary per geospatial region) that is not allowed to be violated during coarsening.
If, at any point, coarsening a family would result in an absolute error greater than the given bound, we do not coarsen.
In order to keep track of the introduced absolute error, it is associated with each resulting element of the coarsening. 

During the first compression step the introduced absolute error, resulting from replacing the initial values $x_i$ that correspond to the elements $e_{i}$, with $i\in I$, which are considered by the potential coarsening step, with their interpolated value $\Bar{x}\, := \, \coarsenmapplain\left(\left\{x_i | i \in I\right\}\right)$, is calculated and checked for compliance by
\begin{equation}
\varepsilon_{\Bar{e}} := \max_{i\in I}\left\{\left|\Bar{x} - x_{i} \right|\right\} \stackrel{?}{\leq} \varepsilon_{max}.
\label{StdAbsErrorEQ}
\end{equation}

If the maximum absolute error $\varepsilon_{\Bar{e}}$ complies with $\varepsilon_{max}$, we will apply the coarsening and associate the error with the resulting coarser element $\Bar{e}$.

For every further compression iteration, the newly introduced absolute error is added to the previous ones in order to obtain an estimation of the deviation from the initial data to the interpolated data $\Bar{\Bar{x}}\, := \, \coarsenmapplain\left(\left\{\bar{x}_i | i \in \Bar{I}\right\}\right)$.

Let $x_{j}$, $j\in I_{i}$ be the values of the family of elements that have been previously coarsened to the values $\Bar{x}_i$ which are associated to the elements $\Bar{e}_{i}$ with $i\in\Bar{I}$.
We are able to estimate the absolute error per element this coarsening would impose by
\begin{equation}
    \begin{split}
        \left|\Bar{\Bar{x}} - x_{j}\right| &\,\leq \,\, \left|\Bar{\Bar{x}} - \Bar{x}_{i}\right| \, +\, \quad \left|\Bar{x}_{i} - x_{j}\right| \\%,\qquad\qquad j\in I_{i},\, i\in\Bar{I}\\
         & \, \leq \,\, \left|\Bar{\Bar{x}} - \Bar{x}_{i} \right| \, + \, \underbrace{\max_{j\in I_{i}}\left| \Bar{x}_{i} - x_{j}\right|}{} \\
         & \, \leq \,\, \left|\Bar{\Bar{x}} - \Bar{x}_{i} \right| \, + \, \qquad \, \varepsilon_{\Bar{e}_i} \, \qquad \, ,j\in I_{i},\, i\in\Bar{I} % =: \varepsilon_{\Bar{e}_i},\,\, i\in \Bar{I}%\stackrel{?}{\leq} \, \varepsilon_{max},\, i\in \Bar{I}
    \end{split}
\end{equation}

The resulting maximum error of this potential coarsening is gathered and checked for compliance with the user-specified maximum error,
\begin{equation}\label{RelErrCheck}
    \varepsilon_{\Bar{\Bar{e}}} := \max_{i\in\Bar{I}}\left\{ \left|\Bar{\Bar{x}} - \Bar{x}_{i} \right| + \varepsilon_{\Bar{e}_i} \right\} \stackrel{?}{\leq} \varepsilon_{max}.
\end{equation}
If the inequality \eqref{RelErrCheck} is satisfied, the coarsening will be applied and the resulting coarse element $\Bar{\Bar{e}}$ will be associated with the new maximum absolute error.

This procedure is performed iteratively until any further coarsening would violate the error bounds.

\subsubsection{Relative error bounds}\label{LabRelErrBounds}

In case of relative error bounds, the user defines a point-wise relative error bound that may vary per geospatial region.
This bound will not be violated during the compression.

Whereas the initial data is present in the first coarsening iteration, it is substituted by the interpolation in any following compression - i.e. coarsening - iterations.
Therefore, the relative error can be precisely calculated during the first coarsening iteration, but needs to be estimated (with regards to the initial data) for any subsequent coarsening iteration in order to reassure controllable losses.

We denote the user-specified maximum relative point-wise error with $\delta_{max}$.
For the first coarsening iteration, we are using \eqref{RelErrorCriterionFirstCoarsening} to check the interpolated value, $\Bar{x}\, := \, \coarsenmapplain\left(\left\{x_i | i \in I\right\}\right)$, for compliance with the point-wise relative error bounds,
\begin{equation}
	\delta_{\Bar{e}} := \max_{i\in I}\left\{\frac{\left| x_{i} - \Bar{x} \right|}{\left| x_{i} \right|}\right\} \,\stackrel{?}{\leq}\, \delta_{max}.
	\label{RelErrorCriterionFirstCoarsening}
\end{equation}

If the family of elements fulfills \eqref{RelErrorCriterionFirstCoarsening}, the coarsening will be applied. 
In any subsequent coarsening iteration, the initial data has already been replaced by ``coarser'' data.
Therefore, we will need to estimate the introduced inaccuracy with regards to the initial data for all following coarsening iterations.

To do so, we store the maximum deviation as an absolute value from the previous coarsening step and associate it with the resulting coarser element $\Bar{e}$, i.e.
\begin{equation}\label{RelErrorAbsDeviation}
\delta_{\Bar{e},abs} := \max_{i\in I}\left\{\left| \Bar{x} - x_{i}\right|\right\}.
\end{equation}

For any further coarsening iteration, we are estimating the point-wise relative error the interpolated value $\Bar{\Bar{x}}\, := \, \coarsenmapplain\left(\left\{\bar{x}_i | i \in \Bar{I}\right\}\right)$ would introduce with respect to the initial data with the help of the previous data values of the considered family $\Bar{x}_{i}$, and the elements' previous absolute maximum inaccuracy $\delta_{\Bar{e}_{i},abs}$, $i\in \Bar{I}$.

Let $x_{j}$, $j\in I_{i}$ be the values of the family of elements that have been previously coarsened to the values $\Bar{x}_i$ which are associated to the elements $\Bar{e}_{i}$ with $i\in\Bar{I}$.
Since, we store the maximum absolute deviation of each coarsening (see \eqref{RelErrorAbsDeviation}), the initial data needs to reside within the interval $x_{j}\in\left[ \Bar{x}_{i} - \delta_{\Bar{e}_{i},abs}, \Bar{x}_{i} + \delta_{\Bar{e}_{i},abs}\right]$ for $j\in I_{i}, i\in\Bar{I}$.

We start the estimation by adapting the enumerator of the relative error,
\begin{equation}\label{RelErrEstimate1}
    \frac{\left| x_{j} - \Bar{\Bar{x}}\right|}{\left| x_{j}\right|} \leq \frac{\left| x_{j} - \Bar{x}_{i}\right| + \left| \Bar{x}_{i} - \Bar{\Bar{x}}\right|}{\left| x_{j}\right|} \leq \frac{\delta_{\Bar{e}_{i},abs} + \left| \Bar{x}_{i} - \Bar{\Bar{x}}\right|}{\left| x_{j}\right|}.
\end{equation}

In order to obtain an estimation of the relative point-wise error we are imposing with this potential coarsening step (with regard to the initial data), we still need to estimate the denominator of \eqref{RelErrEstimate1}.
We are assuming and enforcing that the maximum point-wise relative error is bounded by~$100\%$,
\begin{equation}\label{RelErrorRestriction}
    \delta_{\Bar{e}_{i}} \leq 1.0 \Leftrightarrow \frac{\delta_{\Bar{e}_{i}, abs}}{\left|x_{j}\right|} \leq 1.0 \Leftrightarrow \delta_{\Bar{e}_{i}, abs} \leq \left|x_{j}\right|,\, j\in I_{i},
\end{equation}
which is a reasonable constraint in our application.

Therefore, we are obtaining an estimator for $\left|x_j\right|$ in \eqref{RelErrEstimate1}
by
\begin{equation}\label{RelErrDenominatorInequality}
    \min\left\{ \left|\Bar{x}_{i} -\delta_{\Bar{e}_{i},abs}\right|; \left|\Bar{x}_{i}\right|; \left|\Bar{x}_{i} +\delta_{\Bar{e}_{i},abs}\right|\right\}\, \leq\, \left|x_{j}\right|
\end{equation}
which we proof by cases in the following.

Since we have no access to the initial data anymore, we are regarding all possible cases for the value of $x_{j}$ and show that \eqref{RelErrDenominatorInequality} holds when we assume \eqref{RelErrorRestriction}.

First, let us check the case that $\Bar{x}_{i} \geq x_{j} > 0$.
It holds that,
\begin{equation}
    \begin{split}
        x_{j} \geq \Bar{x}_{i} - \delta_{\Bar{e}_{i},abs} &\Rightarrow \underbrace{x_{j}}_{>0} \geq \underbrace{\Bar{x}_{i} - \delta_{\Bar{e}_{i},abs}}_{\geq 0}, \text{ for } \Bar{x}_{i} \geq \delta_{\Bar{e}_{i},abs} \\
         &\Rightarrow \left| x_{j}\right| \geq \left|x_{i}-\delta_{\Bar{e}_{i},abs}\right|, \text{ for } \Bar{x}_{i} \geq \delta_{\Bar{e}_{i},abs}.
    \end{split}
\end{equation}
In this case, $\Bar{x}_{i} < \delta_{\Bar{e}_{i},abs}$ would indicate that a coarsening has been applied contradicting to our 
constraint \eqref{RelErrorRestriction}, since $0<x_{j}\leq \Bar{x}_{i} < \delta_{\Bar{e}_{i},abs} \nless \left|x_{j}\right|$.

Regarding $\Bar{x}_{i} \geq 0$ and $x_{j} < 0$ leads to a contradiction of \eqref{RelErrorRestriction} as well if $\Bar{x}_{i} > 0$ due to
\begin{equation}
    \delta_{\Bar{e}_{i},abs} := \max_{\tilde{j}\in I_{i}}\left| \Bar{x}_{i} - x_{\tilde{j}}\right| \geq \left|\Bar{x}_{i}-x_{j}\right| \Rightarrow \delta_{\Bar{e}_{i},abs} \geq \underbrace{\Bar{x}_{i}}_{\geq 0} + \left|x_{j}\right|,
\end{equation}
fulfilling the constraint only for $\Bar{x}_{i} = 0$ which corresponds to a relative error of $100\%$. 

In case the values comply to $0>\Bar{x}_{i}\geq x_{j}$, we trivially obtain that $\left|\Bar{x}_{i}\right| \leq \left|x_{j}\right|$.
Similarly, in case of $x_{j}\geq\Bar{x}_{i}>0$, it follows that $\left|\Bar{x}_{i}\right| \leq \left|x_{j}\right|$.

One of the two leftover cases considers $x_{j}>0$ and $\Bar{x}_{i}\leq 0$ which contradicts with the assumption, because it holds that
\begin{equation}
    \underbrace{x_{j}}_{> 0} \leq \underbrace{\Bar{x}_{i}}_{\leq 0} + \,\delta_{\Bar{e}_{i},abs} \Leftrightarrow \underbrace{x_{j}}_{> 0} + \underbrace{\left|\Bar{x}_{i}\right|}_{\geq 0} \leq \delta_{\Bar{e}_{i},abs}
\end{equation}
yielding that $x_{j} + \left|\Bar{x}_{i}\right|\leq \delta_{\Bar{e}_{i},abs} \stackrel{!}{\leq} \left|x_{j}\right|$, which is only fulfilled for $\Bar{x}_{i}=0$.

Finally, we are regarding values that comply to $0>x_{j}\geq \bar{x}_{i}$.
From the upper bound of $x_{j}$ we receive
\begin{equation}
    \underbrace{x_{j}}_{<0} \leq \underbrace{\Bar{x}_{i}}_{\leq x_{j}} +\,\delta_{\Bar{e}_{i},abs} \stackrel{\delta_{\Bar{e}_{i},abs}\leq \left|\Bar{x}_{i}\right|}{\Rightarrow} \underbrace{x_{j}}_{<0} \leq \underbrace{\Bar{x}_{i}+\delta_{\Bar{e}_{i},abs}}_{\leq 0} \Rightarrow \left|x_{j}\right| \geq \left|\Bar{x}_{i} +\delta_{\Bar{e}_{i},abs}\right|.
\end{equation}

In case it should hold that $\delta_{\Bar{e}_{i},abs}>\left|\Bar{x}_{i}\right|$, we obtain
\begin{equation}
    \delta_{\Bar{e}_{i},abs} := \max_{\tilde{j}}\left|\Bar{x}_{i} - x_{\tilde{j}}\right| \geq \underbrace{\left|\Bar{x}_{i}\right.}_{\leq x_{j}}-\underbrace{\left. x_{j}\right|}_{<0} \Rightarrow \left|\Bar{x}_{i}\right| + \left|x_{j}\right| \leq \delta_{\Bar{e}_{i},abs} \stackrel{!}{\leq} \left|x_{j}\right|
\end{equation}
which would only be fulfilled by $\left|\Bar{x}_{i}\right| = 0$ corresponding to a point-wise relative error of $100\%$.

Therefore, we receive an estimator for $\left|x_{j}\right|$ in \eqref{RelErrEstimate1} that does not enlarge the initial absolute value by \eqref{RelErrDenominatorInequality}, yielding
\begin{equation}\label{RelErrEstimate}
    \frac{\left| x_{j} - \Bar{\Bar{x}}\right|}{\left| x_{j}\right|} \leq \frac{\delta_{\Bar{e}_{i},abs} + \left| \Bar{x}_{i} - \Bar{\Bar{x}}\right|}{ \min\left\{ \left|\Bar{x}_{i} -\delta_{\Bar{e}_{i},abs}\right|; \left|\Bar{x}_{i}\right|; \left|\Bar{x}_{i} +\delta_{\Bar{e}_{i},abs}\right|\right\}},\,j\in I_{i},\,i\in\Bar{I},
\end{equation}
which gives us an estimator for the point-wise relative error during the compression iterations,
\begin{equation}
    \delta_{\Bar{\Bar{e}}} := \max_{i\in\Bar{I}}\left\{\frac{\delta_{\Bar{e}_{i},abs} + \left| \Bar{x}_{i} - \Bar{\Bar{x}}\right|}{ \min\left\{ \left|\Bar{x}_{i} -\delta_{\Bar{e}_{i},abs}\right|; \left|\Bar{x}_{i}\right|; \left|\Bar{x}_{i} +\delta_{\Bar{e}_{i},abs}\right|\right\}}\right\}.
\end{equation}

If the estimated point-wise relative error fulfills the user-specified error bound
\begin{equation}
    \delta_{\Bar{\Bar{e}}} \stackrel{?}{\leq} \delta_{max}
\end{equation}
we apply the coarsening and calculate the absolute error it introduces
\begin{equation}
    \delta_{\Bar{\Bar{e}},abs} := \max_{i\in \Bar{I}}\left\{\left| \Bar{\Bar{x}} - \Bar{x}_{i}\right| + \delta_{\Bar{e}_{i},abs}\right\},
\end{equation}
which will then be stored and associated with the resulting coarse element $\Bar{\Bar{e}}$.

\subsection{Flexible Error Criteria}
\begin{figure}\label{DiffErrorDomains}
    \centering
    \includegraphics[width=0.75\textwidth]{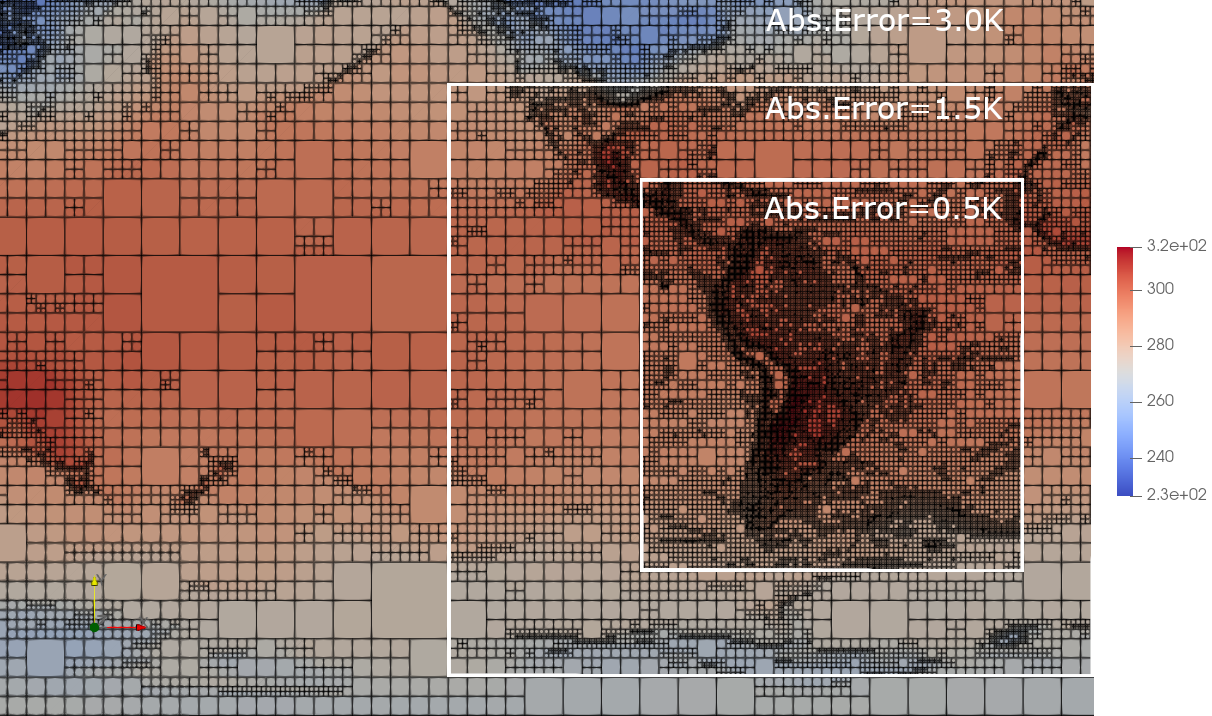}
    \caption{An excerpt from an ERA5 temperature variable (units: Kelvin).
    %defined on a three dimensional domain of cardinality: $\#$Longitude: 1440, $\#$Latitude: 721, $\#$Levels: 37.
    There are several domains with different absolute error bounds which were respected during the compression.}
\end{figure}

The usage of an underlying mesh and the way the inaccuracies are stored and estimated during the compression allows the mixture of relative and absolute error domains to which the compression complies.
Therefore, the approach allows for a fine-grained selection of different domains within the data or even nested domains and their magnitude of permitted error tolerances as shown in figure~\ref{DiffErrorDomains}.

We have shown an absolute and a relative error criterion in the previous subsections.
However, the criteria may be easily applied and extended to user-specified needs, for example excluding some domains completely from the compression or coarsen other areas to the greatest extent.
Possible applications may be to keep a fine resolution over coastal areas, or to coarsen land or ocean domains entirely.
If the data has a temporal dimension, the criteria are of course applicable to the time dimension as well allowing for example to set a focus on a specified time interval.

\subsubsection{Multivariate data}
Many simulations consider more than a single variable. Especially in the domain of the chemistry of Earth's atmosphere, this is common (e.g.~\cite{messydata}).

Potentially, variables behave similar throughout the domain and can be compressed analogously.
In this case, it is possible to define the variables on a single mesh and compress them simultaneously.

\begin{figure}\label{ComparisonComrpessionModes}
    \centering
    \includegraphics[width=0.75\textwidth]{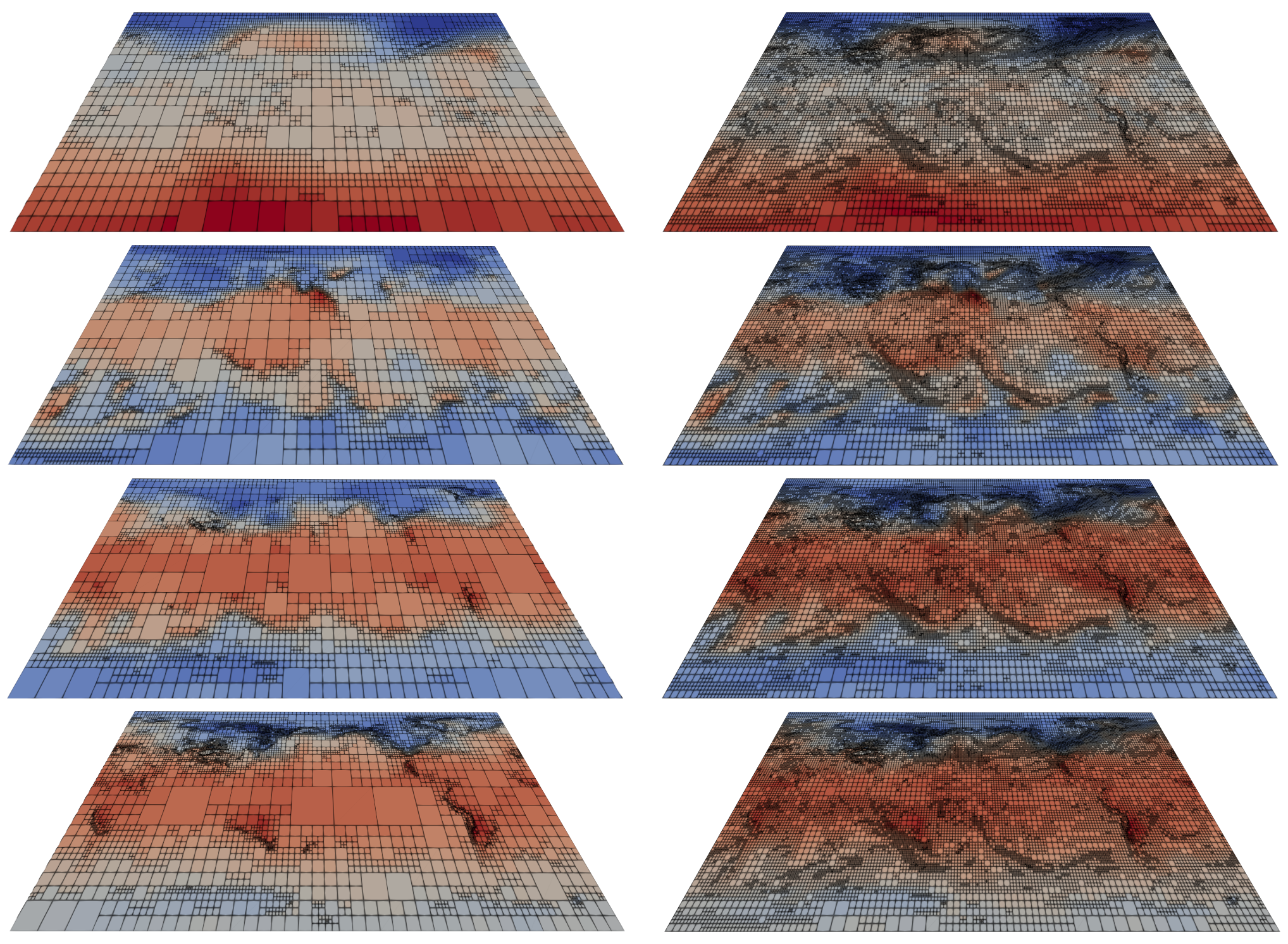}
    \caption{Excerpt of different pressure levels (bottom to top: level $1000$ mbar, $750$ mbar, $250$ mbar, $20$ mbar) at a given time point from an ERA5 temperature variable are shown which resulted from the compression with different modes (left: ``One For One''; right: ``One For All'').
    In order for the different modes to be applicable, the 3D variable has been split beforehand in their ``level'' dimension, resulting in $37$ 2D variables which were compressed independently (``One For One'') and dependently (``One For All'') of each other. In both cases the same relative error criterion has been applied with a relative error bound of $2.5$ percent.}
\end{figure}

We denote this ``compression mode'' as a ``One for All'' (one mesh for all variables) approach.
If the variables are not to be compressed on the same mesh, but rather every variable on it's own, we are labelling this case as a ``One for One'' (one mesh for one variable) approach.
A comparison of both compression modes is shown in figure~\ref{ComparisonComrpessionModes}.

Beneficial in a \textit{One for All} case is that only one mesh has to be managed at a time.
Especially when the compressed data is archived, the introduced overhead for storing the coarsened mesh information is minimized.

However, introducing a coarsening becomes way more restrictive since the adaptation/coarsening criterion needs to be fulfilled for all variables in a certain location of the mesh.

Moreover, our experiments have shown that the data which needs to be stored in order to reconstruct a single mesh is negligible compared to the actual compressed variable's data.

\subsection{Decompression}
Apart from the storage scheme of the compressed data or additional decompression steps due to the combination with supplementary compression techniques, the decompression of the lossy \ac{amr} data is straight-forward once the ``compressed forest'' has been reconstructed (which is described later within this section). 

\begin{figure}
	\begin{center}
		\begin{tikzpicture}[scale=0.5]
				
			% Data points from the compressed data
		\foreach \x in {-10,-9,-6,-5}{
			\foreach \y in {4.5,5.5} {
				\fill[CustomLight] (\x, \y) circle (0.1cm);
			}
		}
	    \fill[CustomDark] (-8.5, 2) circle (0.1cm);
	    \fill[CustomDark] (-7.5, 5) circle (0.1cm);
	    \fill[CustomDark] (-5.5, 3) circle (0.1cm);
	    \fill[CustomLight] (-6, 0.5) circle (0.1cm);
	    \fill[CustomLight] (-5, 0.5) circle (0.1cm);
	    \fill[CustomLight] (-6, 1.5) circle (0.1cm);
	    \fill[CustomLight] (-5, 1.5) circle (0.1cm);
	    
		% Area of dummy elements
		\draw[pattern=north west lines, opacity=0.3] (-10.5,6) -- (-10.5,8) -- (-2.5,8) -- (-2.5,6) --cycle;
		\draw[pattern=north west lines, opacity=0.3] (-4.5,0) -- (-4.5,6) -- (-2.5,6) -- (-2.5,0) --cycle;
		
		% Coarsened/Compressed Mesh
		\draw[black, thin] (-10.5,0) -- (-2.5,0) -- (-2.5,8) -- (-10.5,8) -- cycle;
		\draw[black, thin] (-8.5,8) -- (-8.5,6);
		\draw[black, thin] (-6.5,8) -- (-6.5,6);
		\draw[black, thin] (-4.5,8) -- (-4.5,6);
		\draw[black, thin] (-4.5,6) -- (-2.5,6);
		\draw[black, thin] (-4.5,4) -- (-2.5,4);
		\draw[black, thin] (-4.5,2) -- (-2.5,2);
		\draw[black, ultra thick] (-10.5,0) -- (-4.5,0) -- (-4.5,6) -- (-10.5,6) -- cycle;
		\draw[black, ultra thick] (-10.5,0) -- (-6.5,0) -- (-6.5,4) -- (-10.5,4) -- cycle;
		\draw[black, ultra thick] (-4.5,2) -- (-6.5,2) -- (-6.5,4) -- (-4.5,4) -- cycle;
		\draw[black, ultra thick] (-8.5,4) -- (-6.5,4) -- (-6.5,6) -- (-8.5,6) -- cycle;
		\draw[black, ultra thick] (-6.5,1) -- (-4.5,1);
		\draw[black, ultra thick] (-6.5,5) -- (-4.5,5);
		\draw[black, ultra thick] (-10.5,5) -- (-8.5,5);
		\draw[black, ultra thick] (-9.5,6) -- (-9.5,4);
		\draw[black, ultra thick] (-5.5,6) -- (-5.5,4);
		\draw[black, ultra thick] (-5.5,2) -- (-5.5,0);
		
		% Arrow indicating the mapping
		\draw[black,thick,->] (-2,4) -- (-0.5,4);
		
        % Area of dummy elements
		\draw[pattern=north west lines, opacity=0.3] (0,6) -- (0,8) -- (8,8) -- (8,6) --cycle;
		\draw[pattern=north west lines, opacity=0.3] (6,0) -- (6,6) -- (8,6) -- (8,0) --cycle;
		
		% The grid
		\draw[step=1.0,black,thin] (0,5) grid (6,6);
		\draw[step=2.0,black,thin] (0,6) grid (8,8);
		\draw[step=2.0,black,thin] (6,0) grid (8,6);
		\draw[step=1.0,black, ultra thick] (0,0) grid (6,6);
		\draw[black, ultra thick] (0,0) -- (6,0) -- (6,6) -- (0,6) -- cycle;
		
		% Data points shown as element midpoints
		\foreach \x in {0.5,...,5.5}{
			\foreach \y in {0.5,...,5.5} {
				\fill[CustomLight] (\x, \y) circle (0.1cm);
			}
		}
		\foreach \x in {0.5,...,3.5}{
			\foreach \y in {0.5,...,3.5} {
				\fill[CustomDark] (\x, \y) circle (0.1cm);
			}
		}
		\foreach \x in {2.5,3.5}{
			\foreach \y in {4.5,5.5} {
				\fill[CustomDark] (\x, \y) circle (0.1cm);
			}
		}
		\foreach \x in {4.5,5.5}{
			\foreach \y in {2.5,3.5} {
				\fill[CustomDark] (\x, \y) circle (0.1cm);
			}
		}
		% Memory Layouts 
		\draw[help lines, thick, draw=black] (-4,-1.0) grid[step=0.5] (7,-1.5);
		\draw[help lines, thick, draw=black] (-12,-3) grid[step=0.5] (9.5,-3.5);
		
		%Compressed Layout
		\fill[CustomDark] (-3.75, -1.25) circle (0.1cm);
		\foreach \x in {-3.25,-2.75,...,-1.75} {
		    \fill[CustomLight] (\x, -1.25) circle (0.1cm);
		}
		\draw[pattern=north west lines, opacity=0.3] (-1.5,-1.5) -- (-1,-1.5) -- (-1,-1) -- (-1.5,-1) --cycle;
		\fill[CustomDark] (-0.75, -1.25) circle (0.1cm);
		\draw[pattern=north west lines, opacity=0.3] (-0.5,-1.5) -- (0,-1.5) -- (0,-1) -- (-0.5,-1) --cycle;
		\foreach \x in {0.25,0.75,...,1.75} {
		    \fill[CustomLight] (\x, -1.25) circle (0.1cm);
		}
		\fill[CustomDark] (2.25, -1.25) circle (0.1cm);
		\draw[pattern=north west lines, opacity=0.3] (2.5,-1.5) -- (3.5,-1.5) -- (3.5,-1) -- (2.5,-1) --cycle;
		\foreach \x in {3.75,4.25,...,5.25} {
		    \fill[CustomLight] (\x, -1.25) circle (0.1cm);
		}
		\draw[pattern=north west lines, opacity=0.3] (5.5,-1.5) -- (7,-1.5) -- (7,-1) -- (5.5,-1) --cycle;
		
		%Decompressed Layout
		\foreach \x in {-11.75,-11.25,...,-4.25} {
		    \fill[CustomDark] (\x, -3.25) circle (0.1cm);
		}
		\foreach \x in {-3.75,-3.25,...,-2.25} {
		    \fill[CustomLight] (\x, -3.25) circle (0.1cm);
		}
		\draw[pattern=north west lines, opacity=0.3] (-2,-3) -- (-1.5,-3) -- (-1.5,-3.5) -- (-2,-3.5) --cycle;
		\foreach \x in {-1.25,-0.75,...,0.25} {
		    \fill[CustomDark] (\x, -3.25) circle (0.1cm);
		}
		\draw[pattern=north west lines, opacity=0.3] (0.5,-3) -- (1,-3) -- (1,-3.5) -- (0.5,-3.5) --cycle;
		\foreach \x in {1.25,1.75,...,2.75} {
		    \fill[CustomLight] (\x, -3.25) circle (0.1cm);
		}
		\foreach \x in {3.25,3.75,...,4.75} {
		    \fill[CustomDark] (\x, -3.25) circle (0.1cm);
		}
		\draw[pattern=north west lines, opacity=0.3] (5,-3) -- (6,-3) -- (6,-3.5) -- (5,-3.5) --cycle;
		\foreach \x in {6.25,6.75,...,7.75} {
		    \fill[CustomLight] (\x, -3.25) circle (0.1cm);
		}
		\draw[pattern=north west lines, opacity=0.3] (8,-3) -- (9.5,-3) -- (9.5,-3.5) -- (8,-3.5) --cycle;
		
		% Arrows between compressed and decompressed data
		\draw[CustomDark,->] (-3.75,-1.5) to[out=-150,in=90] (-11.75,-3);
		\draw[CustomDark,->] (-3.75,-1.5) to[out=-145,in=90] (-11.25,-3);
		\draw[CustomDark,->] (-3.75,-1.5) to[out=-140,in=90] (-10.75,-3);
		\draw[CustomDark,->] (-3.75,-1.5) to[out=-135,in=90] (-10.25,-3);
		\draw[CustomDark,->] (-3.75,-1.5) to[out=-130,in=90] (-9.75,-3);
		\draw[CustomDark,->] (-3.75,-1.5) to[out=-125,in=90] (-9.25,-3);
		\draw[CustomDark,->] (-3.75,-1.5) to[out=-120,in=90] (-8.75,-3);
		\draw[CustomDark,->] (-3.75,-1.5) to[out=-115,in=90] (-8.25,-3);
		\draw[CustomDark,->] (-3.75,-1.5) to[out=-110,in=90] (-7.75,-3);
		\draw[CustomDark,->] (-3.75,-1.5) to[out=-105,in=90] (-7.25,-3);
		\draw[CustomDark,->] (-3.75,-1.5) to[out=-100,in=90] (-6.75,-3);
		\draw[CustomDark,->] (-3.75,-1.5) to[out=-95,in=90] (-6.25,-3);
		\draw[CustomDark,->] (-3.75,-1.5) to[out=-95,in=90] (-5.75,-3);
		\draw[CustomDark,->] (-3.75,-1.5) to[out=-90,in=90] (-5.25,-3);
		\draw[CustomDark,->] (-3.75,-1.5) to[out=-90,in=90] (-4.75,-3);
		\draw[CustomDark,->] (-3.75,-1.5) to[out=-90,in=90] (-4.25,-3);
		
		\draw[CustomLight,->] (-3.25,-1.5) to[out=-90,in=90] (-3.75,-3);
		\draw[CustomLight,->] (-2.75,-1.5) to[out=-90,in=90] (-3.25,-3);
		\draw[CustomLight,->] (-2.25,-1.5) to[out=-90,in=90] (-2.75,-3);
		\draw[CustomLight,->] (-1.75,-1.5) to[out=-90,in=90] (-2.25,-3);
		
		% one dummy element
		\draw[CustomLight,dashed,->] (-1.25,-1.5) to[out=-90,in=90] (-1.75,-3);
		
		\draw[CustomDark,->] (-0.75,-1.5) to[out=-90,in=90] (-1.25,-3);
		\draw[CustomDark,->] (-0.75,-1.5) to[out=-90,in=90] (-0.75,-3);
		\draw[CustomDark,->] (-0.75,-1.5) to[out=-90,in=90] (-0.25,-3);
		\draw[CustomDark,->] (-0.75,-1.5) to[out=-90,in=90] (0.25,-3);
		
		% one dummy element
		\draw[CustomLight,dashed,->] (-0.25,-1.5) to[out=-90,in=90] (0.75,-3);
		
		\draw[CustomLight,->] (0.25,-1.5) to[out=-90,in=90] (1.25,-3);
		\draw[CustomLight,->] (0.75,-1.5) to[out=-90,in=90] (1.75,-3);
		\draw[CustomLight,->] (1.25,-1.5) to[out=-90,in=90] (2.25,-3);
		\draw[CustomLight,->] (1.75,-1.5) to[out=-90,in=90] (2.75,-3);
		
		\draw[CustomDark,->] (2.25,-1.5) to[out=-90,in=90] (3.25,-3);
		\draw[CustomDark,->] (2.25,-1.5) to[out=-90,in=90] (3.75,-3);
		\draw[CustomDark,->] (2.25,-1.5) to[out=-90,in=90] (4.25,-3);
		\draw[CustomDark,->] (2.25,-1.5) to[out=-90,in=90] (4.75,-3);
		
		%two dummy elements
		\draw[CustomLight,dashed,->] (2.75,-1.5) to[out=-75,in=90] (5.25,-3);
		\draw[CustomLight,dashed,->] (3.25,-1.5) to[out=-75,in=90] (5.75,-3);
		
		\draw[CustomLight,->] (3.75,-1.5) to[out=-75,in=90] (6.25,-3);
		\draw[CustomLight,->] (4.25,-1.5) to[out=-75,in=90] (6.75,-3);
		\draw[CustomLight,->] (4.75,-1.5) to[out=-75,in=90] (7.25,-3);
		\draw[CustomLight,->] (5.25,-1.5) to[out=-75,in=90] (7.75,-3);
		
		%three dummy elements
		\draw[CustomLight,dashed,->] (5.75,-1.5) to[out=-75,in=90] (8.25,-3);
		\draw[CustomLight,dashed,->] (6.25,-1.5) to[out=-75,in=90] (8.75,-3);
		\draw[CustomLight,dashed,->] (6.75,-1.5) to[out=-75,in=90] (9.25,-3);
		
		\end{tikzpicture}
		\caption{A decompression example. The reconstructed ``compressed forest'' (left) is refined to the initial refinement level of the original data (right). The array containing the compressed data and how a constant interpolation constructs the decompressed data is shown at the bottom. The arrows indicate the copy/insertion-operations corresponding to the constant interpolation.}
		\label{DecompressionExample}
	\end{center}
\end{figure}
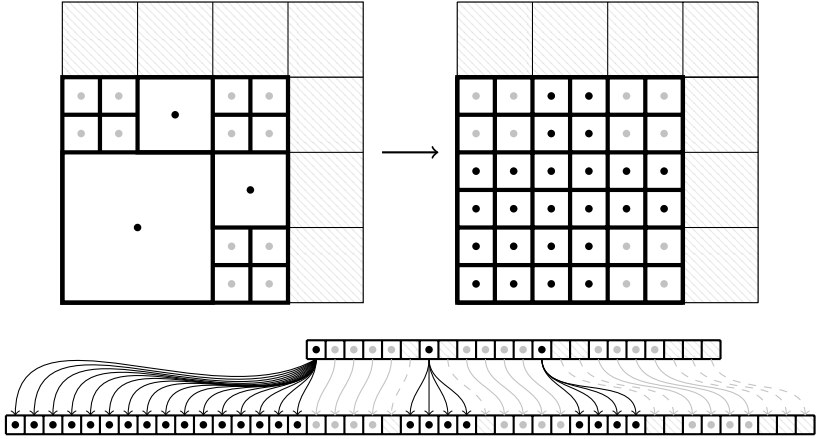

Due to the tree-based \ac{amr} approach and the \ac{sfc}'s implied ordering of the data, the decompression amounts to a plain copy/insertion operation - proportional to the difference between the compressed element's refinement level and the initial refinement level (before the compression); see figure~\ref{DecompressionExample}.

During the decompression, ``dummy elements'' within the compressed forest - laying outside the actual data domain - are easily detected and can be excluded from the decompressed output array.

Afterwards, the data potentially needs to be re-ordered to obtain the initial layout of the input data.

This decompression scheme corresponds to a constant interpolation from the compressed data.

Depending on the magnitude of the permitted error tolerance, this may cause ``blocks'' within the decompressed data.
However, this ensures that the user-defined error tolerance is not violated during the decompression.

Integrating other interpolation methods or smoothing steps during the decompression will introduce additional costs in order to keep track of the allowed error tolerance in order to not violate the given bounds.

\begin{figure}
    \centering
    \includegraphics[width=\textwidth]{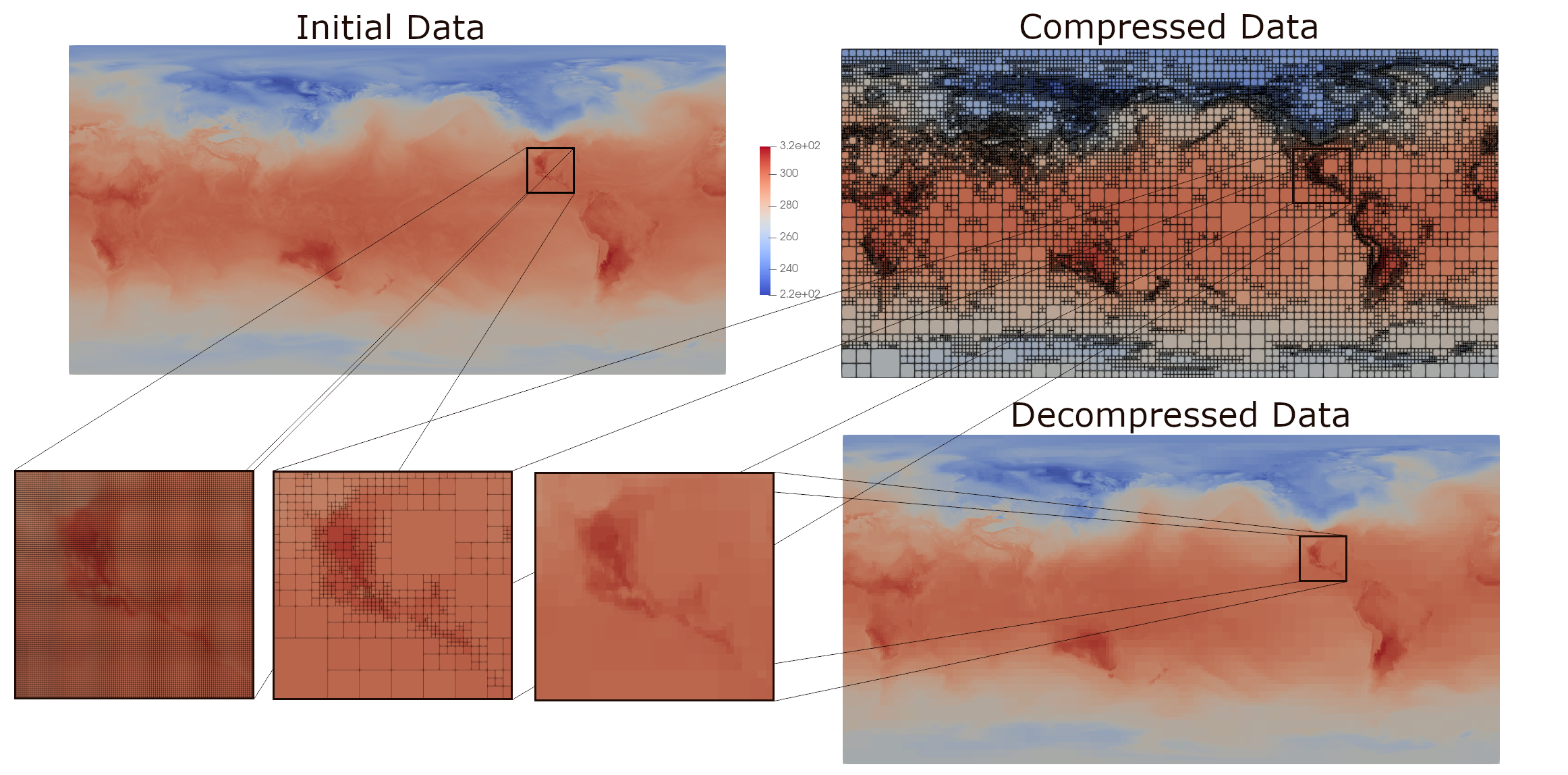}
    \caption{Compression of the $1000$ mbar level (at a fixed point in time) of a three dimensional ERA5 temperature variable (units: \textit{Kelvin}) and it's decompressed data. An absolute error criterion with a maximum permitted error of $2.5$K was utilized.}
    \label{ExampleTemperatureCompressionWithAbsoluteError}
\end{figure}

An exemplary decompression result of an ERA5 3D temperature variable is shown in figure~\ref{ExampleTemperatureCompressionWithAbsoluteError}.
A more in-depth description of this example is given in the following chapter \ref{SecResults}.

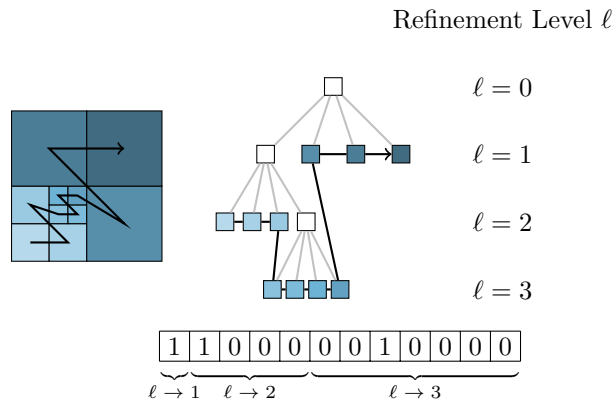
\begin{figure}
    \centering
    \begin{minipage}{0.2\textwidth}
    \vspace*{0.85cm}
    
    \begin{tikzpicture}[scale=0.5]

        \draw[fill=CustomBlueElem1] (-10.5,4) -- (-9.5,4) -- (-9.5,5) -- (-10.5,5) --cycle;
        \draw[fill=CustomBlueElem2] (-9.5,4) -- (-8.5,4) -- (-8.5,5) -- (-9.5,5) --cycle;
        \draw[fill=CustomBlueElem3] (-10.5,5) -- (-9.5,5) -- (-9.5,6) -- (-10.5,6) --cycle;
        \draw[fill=CustomBlueElem4] (-9.5,5) -- (-9,5) -- (-9,5.5) -- (-9.5,5.5) --cycle;
        \draw[fill=CustomBlueElem5] (-9,5) -- (-8.5,5) -- (-8.5,5.5) -- (-9,5.5) --cycle;
        \draw[fill=CustomBlueElem6] (-9.5,5.5) -- (-9,5.5) -- (-9,6) -- (-9.5,6) --cycle;
        \draw[fill=CustomBlueElem7] (-9,5.5) -- (-8.5,5.5) -- (-8.5,6) -- (-9,6) --cycle;
        \draw[fill=CustomBlueElem8] (-8.5,4) -- (-6.5,4) -- (-6.5,6) -- (-8.5,6) --cycle;
        \draw[fill=CustomBlueElem9] (-10.5,6) -- (-8.5,6) -- (-8.5,8) -- (-10.5,8) --cycle;
        \draw[fill=CustomBlueElem10] (-8.5,6) -- (-6.5,6) -- (-6.5,8) -- (-8.5,8) --cycle;
		
		%SFC
        \draw[thick, CustomDark, ->] (-10,4.5) -- (-9,4.5) -- (-10,5.5) -- (-9.25,5.25) -- (-8.75,5.25) -- (-9.25,5.75) -- (-8.75,5.75) -- (-7.5,5) -- (-9.5,7) -- (-7.5,7);
        
    \end{tikzpicture}
    \end{minipage}
    \begin{minipage}{0.45\textwidth}
    \begin{tikzpicture}[scale=0.3]
            %%Tree elements
            %Ebene 0
            \node (celem02) at (5,10) [rectangle, minimum width=0.2cm,draw]{};
            
            %Ebene 1
            \node (celem14) at (2,7) [rectangle, minimum width=0.2cm, draw]{};
            \node (celem15) at (4,7) [rectangle, minimum width=0.2cm, fill=CustomBlueElem8, draw]{};
            \node (celem16) at (6,7) [rectangle, minimum width=0.2cm, fill=CustomBlueElem9, draw]{};
            \node (celem17) at (8,7) [rectangle, minimum width=0.2cm, fill=CustomBlueElem10, draw]{};
            
            %Ebene 2
            \node (celem24) at (0.2,4) [rectangle, minimum width=0.2cm, fill=CustomBlueElem1, draw]{};
            \node (celem25) at (1.4,4) [rectangle, minimum width=0.2cm, fill=CustomBlueElem2, draw]{};
            \node (celem26) at (2.6,4) [rectangle, minimum width=0.2cm, fill=CustomBlueElem3, draw]{};
            \node (celem27) at (3.8,4) [rectangle, minimum width=0.2cm, draw]{};

            %Ebene 3
            \node (celem30) at (2.3,1) [rectangle, minimum width=0.2cm, fill=CustomBlueElem4, draw]{};
            \node (celem31) at (3.3,1) [rectangle, minimum width=0.2cm, fill=CustomBlueElem5, draw]{};
            \node (celem32) at (4.3,1) [rectangle, minimum width=0.2cm, fill=CustomBlueElem6, draw]{};
            \node (celem33) at (5.3,1) [rectangle, minimum width=0.2cm, fill=CustomBlueElem7, draw]{};
            
            \node (reflvltext) at (12.5,13) [] {Refinement Level $\ell$};
            \node (reflvl1) at (12.5,10) [] {$\ell = 0$};
            \node (reflvl2) at (12.5,7) [] {$\ell = 1$};
            \node (reflvl3) at (12.5,4) [] {$\ell = 2$};
            \node (reflvl4) at (12.5,1) [] {$\ell = 3$};

            %% Tree branches
            %Ebene 0-1
            \draw[CustomLight, thick] (celem02) -- (celem14);
            \draw[CustomLight, thick] (celem02) -- (celem15);
            \draw[CustomLight, thick] (celem02) -- (celem16);
            \draw[CustomLight, thick] (celem02) -- (celem17);

            %Ebene 1-2
            \draw[CustomLight, thick] (celem14) -- (celem24);
            \draw[CustomLight, thick] (celem14) -- (celem25);
            \draw[CustomLight, thick] (celem14) -- (celem26);
            \draw[CustomLight, thick] (celem14) -- (celem27);
            
            %Ebene 2-3
            \draw[CustomLight, thick] (celem27) -- (celem30);
            \draw[CustomLight, thick] (celem27) -- (celem31);
            \draw[CustomLight, thick] (celem27) -- (celem32);
            \draw[CustomLight, thick] (celem27) -- (celem33);
            
             %SFC
            \draw[thick, ->, CustomDark] (celem24) -- (celem25) -- (celem26) -- (celem30) -- (celem31) -- (celem32) -- (celem33) -- (celem15) -- (celem16) -- (celem17);

    \end{tikzpicture}
    \end{minipage}
    
    \vspace*{0.25cm}
    
    \begin{tikzpicture}[scale=0.4]
        \node at (0.5,-1) [] {\scriptsize $\ell \rightarrow 1$};
        \node at (3,-1) [] {\scriptsize $\ell \rightarrow 2$};
        \node at (8.5,-1) [] {\scriptsize $\ell \rightarrow 3$};
        
        \node at (0.5,-0.5) [] {\tiny $\underbrace{\hspace{0.36cm}}$};
        \node at (3,-0.5) [] {\tiny $\underbrace{\hspace{1.56cm}}$};
        \node at (8.5,-0.5) [] {\tiny $\underbrace{\hspace{2.76cm}}$};
        
        \node at (0.5,0.5) [] {$1$};
        \node at (1.5,0.5) [] {$1$};
        \node at (2.5,0.5) [] {$0$};
        \node at (3.5,0.5) [] {$0$};
        \node at (4.5,0.5) [] {$0$};
        \node at (5.5,0.5) [] {$0$};
        \node at (6.5,0.5) [] {$0$};
        \node at (7.5,0.5) [] {$1$};
        \node at (8.5,0.5) [] {$0$};
        \node at (9.5,0.5) [] {$0$};
        \node at (10.5,0.5) [] {$0$};
        \node at (11.5,0.5) [] {$0$};
        \draw[step=1.0,black,thin] (0,0) grid (12,1);
    \end{tikzpicture}
    
    \caption{A forest mesh based on a single tree (left) and it's tree-structure (middle to right) are shown. The bit-field, encoding the refinement of elements, which is needed in order to reconstruct the (``compressed'') forest is shown below. A ``1'' indicates a refinement of the corresponding element whereas a ``0'' denotes that the corresponding element has already received it's final status within the mesh. These ``refinement-indicating-bits'' are stored level-wise compliant to the \ac{sfc} indexing of  this tree.}
    \label{EncodedStorageForestMesh}
    \end{figure}

In order to reconstruct the forest, we are storing the refinement from the base mesh - a single tree consisting of only one element in our compression approach - up to the compressed mesh as level-wise bit-fields indicating whether the corresponding leaf elements are further refined (see figure \ref{EncodedStorageForestMesh}).
This leads to a bit-filed generally exhibiting long series of $1$'s or $0$'s promoting the usage of additional lossless compression techniques (e.g. LZ77) to further reduce the storage costs.

Applying these ``refinement-bits'' level-wise to the base mesh reconstructs the compressed mesh iteratively.

\section{Results}\label{SecResults}
We compare different lossy compressors on a sample of variables from the ERA5 datasets \cite{ERA5Data, ERA5tco3}.
For the portrayal of the lossy compression and its error criteria, we have chosen a \textit{temperature} variable from the dataset \cite{ERA5Data} and a \textit{total column ozone} variable from the dataset \cite{ERA5tco3}.

The temperature data is supplied as hourly data on a regular latitude-longitude grid on pressure levels.
We have chosen a single snapshot of this variable at a given point in time (namely January 1, 2022, 0:00) yielding data of the dimensionality:  $\#$longitude:~1440, $\#$latitude:~721, $\#$level:~37.

The hourly data of the total column ozone variable is given on a regular latitude-longitude grid resembling the total amount of ozone in a column of air extending from the surface of the Earth to the top of the atmosphere.
For this variable, we have considered the temporal dimension of the data as well.
We have taken 360 consecutive temporal snapshots of this variable starting at January 1, 2022, 0:00, for our examination; yielding data of dimensionality: $\#$longitude:~1440, $\#$latitude:~721, $\#$time:~360.

The data of both variables is supplied as \textit{short} - 2 byte - integer data which needs to be unpacked by multiplying a scaling factor and adding an offset.
We have applied the transformations and stored the data as single-precision - 4 byte - floating point data.

In order to assess the potential of a compression technique based solely on adaptive coarsening, we compared different error-bounded lossy compressors: SZ \cite{di2016fast} (version: 2.1.12), ZFP \cite{lindstrom2014fixed} (version: 1.0.0), and ISABELA \cite{Lakshminarasimhan13} (version: 0.2.1) to the results of our lossy \ac{amr} compressor.
Therefore, we set different absolute and relative error criteria and compared the resulting compressed data sizes.

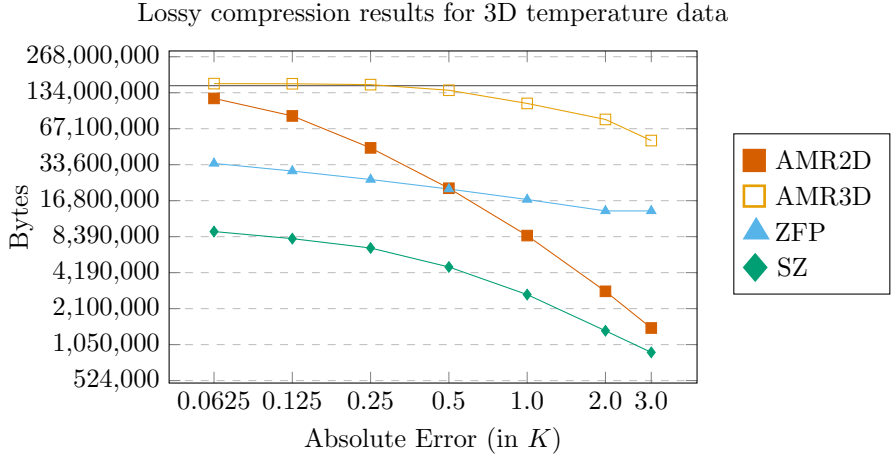
\begin{figure}
		\begin{tikzpicture}[
		    amr2dnode/.style={shape=rectangle,fill=Plot1, inner sep=0pt, minimum width=0.275cm, minimum height=0.275cm, draw=Plot1, line width=1},
            amr3dnode/.style={shape=rectangle, draw=Plot2, inner sep=0pt, minimum width=0.275cm, minimum height=0.275cm, line width=1},
            zfpnode/.style={shape=isosceles triangle, anchor=left corner, inner sep=0pt, minimum width=0.3cm, minimum height=0.25cm, shape border rotate=90, fill=Plot3, draw=Plot3, line width=1,isosceles triangle stretches},
            sznode/.style={shape=diamond, inner sep=0pt, minimum width=0.3cm, minimum height=0.35cm, fill=Plot4, draw=Plot4, line width=1}
		]
			\begin{loglogaxis}[
				title={Lossy compression results for 3D temperature data},
				xlabel={Absolute Error (in $K$)},
				ylabel={Bytes},
				ylabel style={at={(-0.24,0.5)}},
				width=8.6cm,
				height=6cm,
				xmin=0.042, xmax=4.5,
				ymin=500000, ymax=300000000,
				xtick={0.0625,0.125,0.25,0.5,1.0,2.0,3.0},
				xticklabels={$0.0625$,$0.125$,$0.25$,$0.5$,$1.0$,$2.0$,$3.0$},
				log ticks with fixed point,
				log basis y=2,
				ytick={524288,1048576,2097152,4194304,8388608,16777216,33554432,67108864,134217728,268435456},
				legend style={at={(1.02,0.5)},
					fill=white, draw=black,
					anchor=west,legend columns=1},
				ymajorgrids=true,
				grid style=dashed
				]
				\addplot[mark=none, black!70, samples=2,domain=0.02:5.0] {153659520};
				\addplot[
				color=Plot1,
				mark=square*
				]
				coordinates {%Removed coordinate at 0.0 (0.035,153659520)
				    (0.0625,120298704)(0.125,85819182)(0.25,46376586)(0.5,21316265)(1.0,8565727)(2.0,2926041)(3.0,1443839)
				};
				\addplot[
				color=Plot2,
				mark=square,
				]
				coordinates {
					(0.0625,159920758)(0.125,159413467)(0.25,156827808)(0.5,140789474)(1.0,109201123)(2.0,80207578)(3.0,53412234)
				};
				\addplot[
				color=Plot3,
				mark=triangle*
				]
				coordinates {
					(0.0625,34383240)(0.125,29693752)(0.25,25212312)(0.5,21000968)(1.0,17157840)(2.0,13770616)(3.0,13770616)
				};
				\addplot[
				color=Plot4,
				mark=diamond*
				]
				coordinates {
					(0.0625,9254337)(0.125,8073349)(0.25,6740907)(0.5,4676266)(1.0,2747678)(2.0,1368679)(3.0,902879)
				};

			\end{loglogaxis}
			
			\matrix [draw,right] at (7.5,2.25) {
              \node [amr2dnode,label=right:AMR2D] {}; \\
              \node [amr3dnode,label=right:AMR3D] {}; \\
              \node [zfpnode,label=right:ZFP] {}; \\
              \node [sznode,label=right:SZ] {}; \\
            };
            
		\end{tikzpicture}
		\caption{Lossy compression of ERA5 3D temperature data (Dimensionality: 1440 $\times$ 721 $\times$ 37). The resulting byte size of the compressed data is shown in dependency of the permitted absolute error criteria. The raw floating point data made up roughly $154$ MB of storage (the base line of this size is depicted as a thin black line within the figure). Results for the different compressors are displayed in comparison.}\label{TempDataAbsErrorCompr}
	\end{figure}

The unpacked floating point data of the temperature variable makes up roughly $154$ MB of binary data.
In figure~\ref{TempDataAbsErrorCompr} we show the resulting compressed data sizes of the temperature variable utilizing different absolute error criteria.
The comparison consists of the SZ, ZFP and our \ac{amr} compressor.
Since the ISABELA compressor only provides relative error bounds, it had been excluded from this comparison.

We have applied the \ac{amr} compressor as a 3D method as well as a 2D method.
In the latter case - \textit{AMR2D} - the ``level''-dimension of the data has been split resulting in $37$ 2D variables (on a longitude $\times$ latitude domain).
Those $37$ variables have been compressed independently of each other with a ``One For One'' compression mode.

We clearly catch that the application of the 2D methodology yields considerably better compression results in terms of file size than it's 3D counterpart.
Decoupling a dimension and performing the compression on the several lower dimensional data slices potentially removes a high variability in a certain dimension.
Moreover, as shown in figure \ref{CoarseningSchemeQuadHex}, in a 2D case the error criterion only needs to be fulfilled for four elements in order to apply a coarsening, whereas in a 3D case eight elements have to satisfy the error criterion which is more restrictive.
Therefore, a high variability in a certain dimension which is not decoupled deteriorates the compression significantly.

Furthermore, we grasp that the SZ compressor as well as our lossy \ac{amr} compressor yield high data size reduction for increasing permitted absolute errors, therefore, outperforming the ZFP compressor for higher permitted absolute error criteria.

Overall, we recognize a weaker performance for the \ac{amr} compressor for small permitted errors.
This is due to the fact, that the smaller the allowed error is, the less coarsenings will be applicable and since the compressed \ac{amr} data consists of real-valued data (in this case single-precision floating point data), the initial data is stored more or less unaltered plus additional storage for the encoded forest mesh leading to large compressed data sizes.

However, these experiments only examined the applicability of lossy compression solely powered by adaptive coarsening. The combination with additional compression techniques may lead to stronger results.

Considering that the already competitive results for medium to large permitted errors, the \ac{amr} compression exhibits a promising compression approach.

\begin{figure}
		\begin{tikzpicture}[
		    amr2dnode/.style={shape=rectangle,fill=Plot1, inner sep=0pt, minimum width=0.275cm, minimum height=0.275cm, draw=Plot1, line width=1},
            isabelanode/.style={shape=isosceles triangle, anchor=left corner, inner sep=0pt, minimum width=0.3cm, minimum height=0.25cm, shape border rotate=90, fill=Plot3, draw=Plot3, line width=1,isosceles triangle stretches},
            sznode/.style={shape=diamond, inner sep=0pt, minimum width=0.3cm, minimum height=0.35cm, fill=Plot4, draw=Plot4, line width=1}
		]
			\begin{semilogyaxis}[
				title={Lossy compression results for temporal total column ozone data},
				xlabel={Relative Error (in $\%$)},
				ylabel={Bytes},
				ylabel style={at={(-0.28,0.5)}},
				width=8.6cm,
				height=6cm,
				xmin=0.5, xmax=5.5,
				ymin=6000000, ymax=2300000000,
				xtick={1.0,2.0,3.0,4.0,5.0},	xticklabels={$1.0$,$2.0$,$3.0$,$4.0$,$5.0$},
				log ticks with fixed point,
				log basis y=2,	ytick={8192000,16384000,32768000,65536000,131072000,262144000,524288000,1048576000,2097152000},
				ymajorgrids=true,
				grid style=dashed,
				]
				\addplot[mark=none, black!70, samples=2,domain=0.02:6.0] {1495065600};
				\addplot[
				color=Plot1,
				mark=square*,
				]
				coordinates {
					(1.0,123920044)(2.0,44952352)(3.0,23314368)(4.0,14261384)(5.0,9629832)
				};
				\addplot[
				color=Plot3,
				mark=triangle*,
				]
				coordinates {
					(1.0,528802480)(2.0,528225237)(3.0,528179043)(4.0,528172239)(5.0,528170044)
				};
				\addplot[
				color=Plot4,
				mark=diamond*,
				]
				coordinates {
					(1.0,37089312)(2.0,22998802)(3.0,19629064)(4.0,17486285)(5.0,11104400)
				};

				%\legend{AMR2D, ISABELA, SZ}
			\end{semilogyaxis}
			
			\matrix [draw,right] at (7.5,2.25) {
              \node [amr2dnode,label=right:AMR2D] {}; \\
              \node [isabelanode,label=right:ISABELA] {}; \\
              \node [sznode,label=right:SZ] {}; \\
            };
            
            \hspace*{1cm}
		\end{tikzpicture}
		\caption{Lossy compression of temporal ERA5 total column ozone data (Dimensionality: 1440 $\times$ 721 $\times$ 360).
		The raw floating point data made up roughly $1.5$ GB; this initial data size is depicted by a thin black line).
		The resulting compressed file sizes of the data for different relative error criteria are shown in comparison for the SZ, ISABELA and \ac{amr} compressor.
		The \ac{amr} compressor split the time-dimension and performed a ``One for One'' compression for the resulting $360$ latitude-longitude slices; denoted as the AMR2D compressor. 
		}\label{TCO3RelErrorCriterionCompression}
	\end{figure}
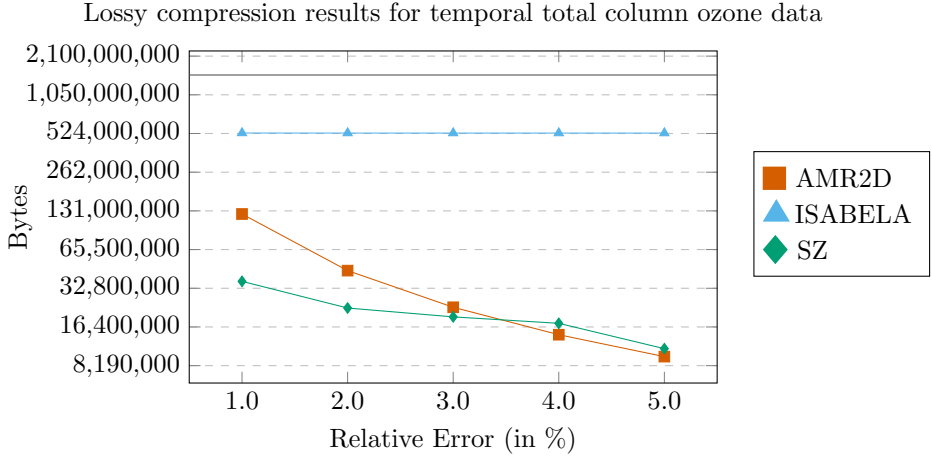

    The (unpacked) floating point data of the considered total column ozone variable makes up nearly $1.5$GB of binary data.
    Within figure \ref{TCO3RelErrorCriterionCompression} we compared the file sizes of the compression results for relative error-bounded lossy compression of the SZ, ISABELA and our AMR2D compressor.
    Since the ZFP compressor does not provide an option for point-wise relative error bounds, it has been excluded from this comparison.
    
    We have compared the results of the SZ and ISABELA compressors with the AMR2D compressor on different relative error thresholds.
    The three-dimensional data has been split in its time-dimension by the AMR compressor resulting in $360$ latitude-longitude-slices which have been compressed independently of each other by the AMR compression.
    At first sight, we notice that the resulting file sizes of the compressed data from the SZ and AMR2D compressor are significantly smaller than the ones originating from the ISABELA compression.
    The ISABELA compressor yield approximately the same file sizes for the different relative error criteria lowering the output only by a few megabytes for higher permitted errors.
    
    Similar to the results for the absolute error criteria, we see that the AMR compression yields weaker results - in terms of the compressed file sizes - than the SZ compressor for small relative errors but becomes more competitive and even outperforms the SZ compressors for relatively high permitted errors.

	\subsection{Packed Data}
	Climate data is often stored and shared in netCDF files.
	A common practice to limit the size of the output data is ``packing'' of the data.
	Essentially, this is done by limiting the precision of the data and storing it in smaller data types, e.g. convert floating point data to short integer (2 byte) data with a limited precision.
	In order to apply this conversion a scale factor and an offset needs to be generated defining the transformation.
	A common scheme for packing the data is shown in \cite{NcPackedData}, such that the data can be regained for example~by
	\begin{equation}
	    \underbrace{unpacked\_value}_{\text{float}} = \underbrace{scale\_factor}_{\text{float}} \cdot \underbrace{packed\_value}_{\text{short integer}} + \underbrace{offset}_{\text{float}}.
	\end{equation}
	
	Since our lossy compression approach is applicable to any type of data, we are able to run the compression directly on the packed data.
	Therefore, we just need to transform the absolute error criterion into the range of the packed data by dividing by the scale factor of this variable,
	\begin{equation}
	    \varepsilon_{\text{packed}} = \frac{\varepsilon_{\text{unpacked}}}{scale\_factor}.
	\end{equation}
	
\begin{figure}
		\begin{tikzpicture}[
		    amr2dnode/.style={shape=rectangle,fill=Plot1, inner sep=0pt, minimum width=0.275cm, minimum height=0.275cm, draw=Plot1, line width=1},
            amr3dnode/.style={shape=rectangle, draw=Plot2, inner sep=0pt, minimum width=0.275cm, minimum height=0.275cm, line width=1},
            zfpnode/.style={shape=isosceles triangle, anchor=left corner, inner sep=0pt, minimum width=0.3cm, minimum height=0.25cm, shape border rotate=90, fill=Plot3, draw=Plot3, line width=1,isosceles triangle stretches},
            sznode/.style={shape=diamond, inner sep=0pt, minimum width=0.3cm, minimum height=0.35cm, fill=Plot4, draw=Plot4, line width=1},
            amr2dpacked/.style={shape=circle, fill=Plot5, draw=Plot5, line width=1, inner sep=0pt, minimum size=0.275cm}
		]
			\begin{loglogaxis}[
				title={Lossy compression results for 3D temperature data},
				xlabel={Absolute Error (in $K$)},
				ylabel={Bytes},
				ylabel style={at={(-0.24,0.5)}},
				width=8.6cm,
				height=6cm,
				xmin=0.042, xmax=4.5,
				ymin=500000, ymax=300000000,
				xtick={0.0625,0.125,0.25,0.5,1.0,2.0,3.0},
				xticklabels={$0.0625$,$0.125$,$0.25$,$0.5$,$1.0$,$2.0$,$3.0$},
				log ticks with fixed point,
				log basis y=2,
				ytick={524288,1048576,2097152,4194304,8388608,16777216,33554432,67108864,134217728,268435456},
				ymajorgrids=true,
				grid style=dashed
				]
				\addplot[mark=none, black!70, samples=2,domain=0.02:5.0] {76829760};
				\addplot[mark=none, black!70, samples=2,domain=0.02:5.0] {153659520};
				\addplot[
				color=Plot1,
				mark=square*
				]
				coordinates {
				    (0.0625,120298704)(0.125,85819182)(0.25,46376586)(0.5,21316265)(1.0,8565727)(2.0,2926041)(3.0,1443839)
				};
				\addplot[
				color=Plot2,
				mark=square
				]
				coordinates {
					(0.0625,159920758)(0.125,159413467)(0.25,156827808)(0.5,140789474)(1.0,109201123)(2.0,80207578)(3.0,53412234)
				};
				\addplot[
				color=Plot3,
				mark=triangle*
				]
				coordinates {
					(0.0625,34383240)(0.125,29693752)(0.25,25212312)(0.5,21000968)(1.0,17157840)(2.0,13770616)(3.0,13770616)
				};
				\addplot[
				color=Plot4,
				mark=diamond*
				]
				coordinates {
					(0.0625,9254337)(0.125,8073349)(0.25,6740907)(0.5,4676266)(1.0,2747678)(2.0,1368679)(3.0,902879)
				};
				\addplot[
				color=Plot5,
				mark=oplus*
				]
				coordinates {
					(0.0625,62397316)(0.125,43476320)(0.25,23618104)(0.5,10860460)(1.0,4380776)(2.0,1507756)(3.0,753096)
				};
			\end{loglogaxis}
			
			%\hspace*{-0.2cm}
			
			\matrix [draw,right] at (7.5,2.25) {
              \node [amr2dnode,label=right:AMR2D] {}; \\
              \node [amr3dnode,label=right:AMR3D] {}; \\
              \node [zfpnode,label=right:ZFP] {}; \\
              \node [sznode,label=right:SZ] {}; \\ \hline[thick]
              \node[amr2dpacked,label=right:AMR2D-Packed]{};\\
            };
		\end{tikzpicture}
		\caption{Lossy compression of ERA5 3D temperature data (Dimensionality: 1440 $\times$ 721 $\times$ 37). The resulting byte size of the compressed data is shown in dependency of the permitted absolute error criteria. The raw floating point data made up roughly $154$ MB of storage and the raw packed short integer data made up roughly $77$ MB (both depicted with a thin black line). The results of the compression of the packed data (denoted as ``AMR2D-Packed'') are shown for different absolute error criteria. The previous results of the floating point compression are displayed for context.}\label{TempDataAbsErrorComprWithPackedData}
	\end{figure}
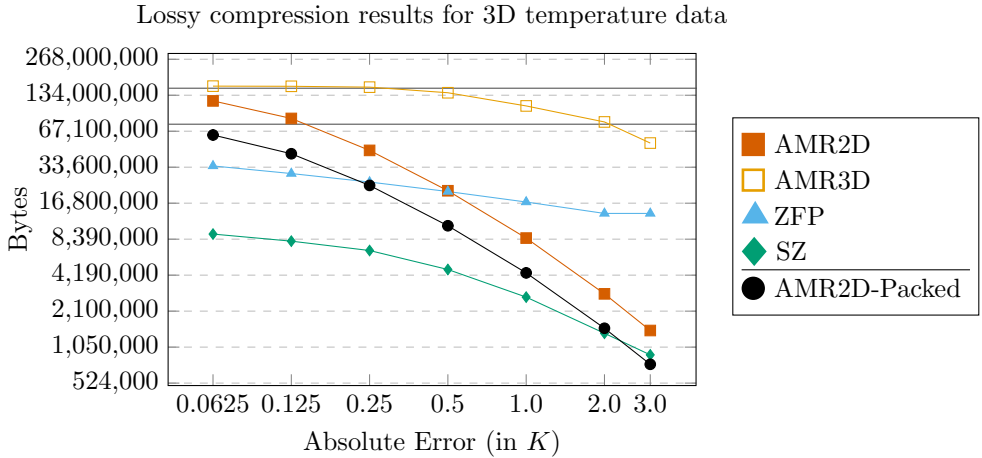

	Regarding the example of the ERA5 temperature data from above, we enhance the lossy AMR compression by nearly dividing the compressed data size in half, since the data to be stored is only half the size of the unpacked data (i.e. 2-byte integer instead of 4-byte floating point data); see figure \ref{TempDataAbsErrorComprWithPackedData}.

\section{Discussion}\label{SecDiscussion}
We have presented a lossy compression approach based on \ac{amr} techniques.
With the given adaptation criteria, it is easily implemented in existing \ac{amr} applications in order to lower the output.
It also applicable as a general compression technique for example for multidimensional data given in the form of arrays or may be applied as a pre-processing step for additional compression methods.

\subsection{Disadvantages}
The lossy compression based on \ac{amr} will most likely never exploit the full magnitude of the permitted error tolerance on it's own and therefore, will not yield a stand-alone compression approach.
This is due to the fact, that a coarsening step (based on a pre-defined coarsening scheme) is either applied or not.
If it is not applied, we will generally still remain with some ``unused'' amount of allowed error tolerance which may be exploited by different compressors.
Moreover, the approach reduces the overall amount of data values.
The method itself makes no effort to further find and reduce redundancies on the byte or even bit-level.

\subsection{Advantages}
Since many applications produce gridded data, it is a natural approach to coarsen the data on the grid in some areas in order to reduce the amount of data.
Furthermore, the application of this approach can be highly parallel.

Due to the strong geospatial dependency of the technique, several domains of interest within the data may be excluded from the compression, or domains with different permitted error bounds can be nested.

Data dimensions exhibiting a high variability within the data may deteriorate the compression.
Therefore, our \ac{amr} compressor is capable of splitting a dimension (for example the \textit{elevation} in a geospatial context) leading to better compression ratios than compressing the whole variable as it is.

Moreover, these techniques are general in nature, i.e., they are applicable to all arithmetic data types (e.g. floating point and integer data).

\section{Conclusion}\label{SecConclusion}
The portrayed compression techniques based on \ac{amr} are especially suitable for geospatial data with a certain local smoothness.
The adaptation criteria from section \ref{LabRelErrBounds} and \ref{LabAbsErrBounds} being used for the compression allow for controllable losses.
The compression has shown to be very flexible in terms of defining domains within the data that are ought to be compressed with different error criteria or splitting up a variable into several lower-dimension variables by decoupling a dimension with high variability.

Being based on tree-based \ac{amr}, the compression approach is highly parallel which makes this method in combination with the aforementioned features especially suitable for large-scale geospatial data sets.

In comparison to current state of the art lossy compressors, our lossy AMR data reduction technique achieves competitive compression results in terms of file size reduction.
The compressor is implemented within our open source tool cmc \cite{CMC}.

It needs to be emphasized that this technique is completely ``value-based'' if considered stand-alone.
In particular, the compressed data still consists of values of the initial data type (e.g. single-precision floating point data).

Therefore, a combination of the lossy \ac{amr} compressor with supplementary compression techniques seems reasonable and promising.
First experiments of pipelining the \ac{amr} compression with current state of the art compression techniques indicate auspicious results which we will examine thoroughly in the future.

The portrayal of the compression techniques within this paper was bound to quadrilateral and hexahedral mesh elements onto which the data has been mapped.
Since our underlying AMR library t8code is capable of handling adaptive meshes with various element types, an extension of the compressor to operate on data defined on different element shapes, for example originating from refined geodesic grids as used in the ICON model, is imaginable and will be conducted in the future.
Instead of utilizing a constant approximation on the elements within the compression scheme, an ansatz of higher polynomial degree may be a promising enhancement, in particular for relatively small permitted error tolerances.

\section*{Acknowledgments}
This work was performed as part of the Helmholtz School for Data Science in Life, Earth and Energy (HDS-LEE) and received funding from the Helmholtz Association of German Research Centres.\\
Gregor Gassner acknowledges funding from the Federal Ministry of Education and Research (BMBF) through the projects ADAPTEX (FKZ 16ME0672) and ICON-DG (01LK2315B) and from the DFG through the research unit SNuBIC (DFG-FOR5409).

\bibliographystyle{siamplain}
\bibliography{references}
\end{document}